# RESOLUTION ANALYSIS OF FILMS WITH EMBEDDED SPHERES FOR IMAGING OF NANOPLASMONIC ARRAYS

by

Navid Farahi

A thesis submitted to the faculty of
The University of North Carolina at Charlotte
in partial fulfillment of the requirements
for the degree of Master of Science in
Optical Science and Engineering

Charlotte

2015

Approved by:

___________________________
Dr. Vasily Astratov

___________________________
Dr. Irina Nesmelova

___________________________
Dr. Yasin Raja

# ABSTRACT

NAVID FARAHI.  Resolution analysis of films with embedded microsphere for Imaging of nanoplasmonic arrays.  (Under the direction of DR. VASILY ASTRATOV)

With the advent of microsphere assisted microscopy in 2011, this technique emerged as a simple and easy way to obtain optical super-resolution. Although the possible mechanisms of imaging by microspheres are debated in the literature, most of the experimental studies established the resolution values well beyond the diffraction limit. It should be noted, however, that there is no standard resolution measurement in this field that researchers can use. The reported resolution has been based on the smallest discernible feature; although it seems logical but it is not based on the standard textbook definition, and so far it has ended to a wide range of resolution reports based on qualitative criteria which can lead to exaggerated resolution values. In addition, this method has another limitation related to its limited field-of-view. In this work, first we fabricated a novel optical component for super-resolution imaging based on an attachable polydimethylsiloxane (PDMS) thin film with embedded high index ($n\sim2$) barium titanate glass (BTG) microspheres. It is shown that such films can be translated along the surface of investigated structures to enhance field-of-view. Second, we introduced a method of image treatment which allows determining the super-resolution values consistent with the resolution definition in the conventional diffraction-limited optics. We demonstrated this method for a typical microsphere-assisted image where we measured the super-resolution of $\sim\lambda/5.5$. We also developed this technique to measure the resolution of a micro-cylindrical-assisted system.

# DEDICATION

*I dedicate this thesis to my...*

*...parents for their endless and unconditional love, support and guidance*

*... wife for her unswerving faith and support*

*... brothers for their support and encouragement*

# ACKNOWLEDGEMENTS

First of all, I would like to express my keen gratitude to my advisor, Dr. Vasily N. Astratov for his continuous support during last 2 years with his absolutely remarkable knowledge and experience. Without his guidance and encouragement the work presented in this thesis could not have been accomplished.

I would also like to thank my committee members, Dr. Yasin Raja and Dr. Irina Nesmelova for accepting to serve on my committee and also their time and comments.

I am grateful to, Dr. Glenn Boreman, Dr. Faramarz Farahi, Dr. Pedram Leilabady, Dr. Angela Davies, Dr. Greg Gbur, Dr. Michael A. Fiddy, Dr. Ed. Stokes, Dr. Yildirim Aktas, Dr. Yuri Nesmelov, Dr. Tsing-Hua Her and Mr. Hudson welch for their precious help, discussions and insights during my courses and academic life.

I am very thankful to Dr. Nicholaos I. Limberopoulos, Dr. Augustine M. Urbas, Dennis E. Walker Jr., and Dr. Dean P. Brown for providing my internship at the Air Force Research Laboratory at Wright-Patterson Air Force Base.
During my graduate studies, I enjoyed my collaboration with my fellow
Mesophotonics Laboratory members, Dr. Yangcheng Li, Farzaneh Abolmaali, Dr. Kenneth W. Allen and Dr. Arash Darafsheh.

I would like to thank fellow graduate students for productive discussions Dr. Mehrdad Abolbashari, Alexander Bermudez, Dr. Hossein Alisafaee, Dr. Gelareh Babaie, Nasim Habibi, Ali Pouya Fard, Dr. Mahsa Farsad, Elisa Hurwitz, Zeba Naqvi, Mark Green, Joseph Peller, Morteza Ghaempanah, Jack Chung, Jinghua Ge, Jonathan Babaie, Michael Uwakwe, Christopher Wilson, Dr. Michele (Yue) Dong, Mohammad Azari, Dr. Jason Case, Frances Bodrucki, Herminso Villarraga-Gomez.

I am grateful to Dr. Lou Deguzman, Dr. Robert Hudgins, and Scott Williams for all their trainings and kind support during my research at UNCC Optics center.

I also thank Mark Clayton, Wendy Ramirez, and Elizabeth Butler for assisting me throughout all administrative affairs.

I am grateful to Dr. Glen Marrs (Wake Forrest University), Dr. Didier Dreau and David Gray (UNC-Charlotte), and Dr. Robert Peterson (Olympus Scientific Solutions Americas) for their kind collaboration, productive discussion and training during my research.

I was supported by the GASP award throughout my graduate studies. I am thankful for the Department of Physics and Optical Sciences for supporting me with teaching assistantships and also grateful for Dr. Vasily N. Astratov's for providing me with research assistantship through his grants from the National Science Foundation.

At last, this work would have not been possible without my family to whom this thesis is dedicated.

# TABLE OF CONTENTS

## LIST OF FIGURES

# CHAPTER 1: INTRODUCTION

## 1.1    Resolution of the Optical System

In 1873, Ernst Karl Abbe published a paper on the resolution of optical microscope [1, 2]. Using wave optics, he showed that there is a limit for conventional optical microscope to resolve an object and this limit was written in the form:

$$d = K \frac{\lambda}{NA} \ , \tag{1}$$

where $d$ is the distance between the two point sources, $\lambda$ is the wavelength of the light used for imaging and $NA$ is numerical aperture which is defined by:

$$NA = nSin\theta \ , \tag{2}$$

where $\theta$ is the half angle of the cone of light which can enter the lens (objective) from a point source, see Figure 1.1, and $n$ is the refractive index of the object-space.

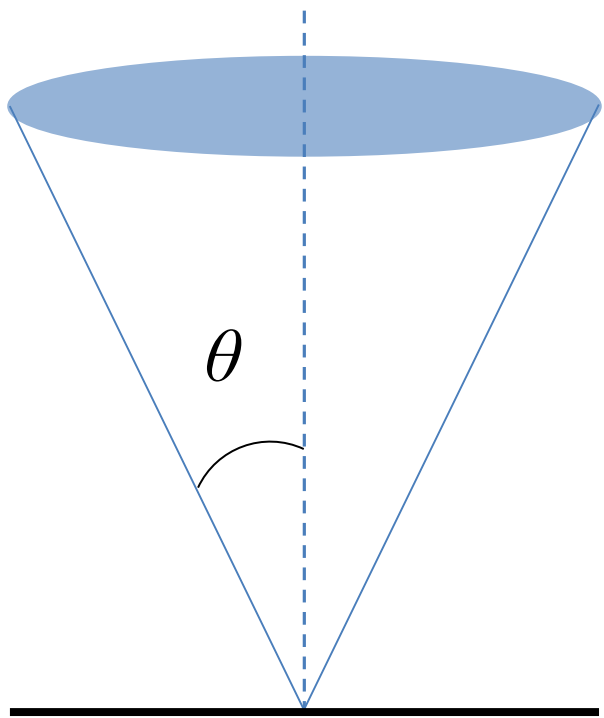

Figure 1.1: Shows a schematic diagram of the cone of light limited by the aperture lens to the half angle θ.

The parameter $K$ in the Eq. (1) represents a somewhat arbitrary criterion of resolution of two point objects. There is a series of definitions of resolution which results in the parameter K varying from 0.473 to 0.61. Rayleigh's criterion [2, 3] is defined as the distance of two point sources where the center of maximum irradiance of one source is on the first minimum of the other. Applying circular aperture (since in current thesis we assume we have a circular aperture) the summation of both irradiance make the final irradiance whose profile has a saddle to pick (S/P) ratio of 73.6% for Rayleigh's criterion [4]. Passing through the circular aperture, the light makes the irradiance pattern to be an Airy disk which also called the point spread function (PSF) of the system [4, 5].

$$I(r) = I_0 (\frac{2J_1(k_0 rNA)}{k_0 rNA})^2, \tag{3}$$

where $J_1$ is the Bessel function of first kind of order 1 and $k_0$ is $\frac{2\pi}{\lambda_0}$ the free-space wave number of the illuminating light. The Rayleigh's criterion consequently corresponds to $K$=0.61 which means that:

$$d = \frac{0.61\lambda}{NA}, \tag{4}$$

The next resolution criterion was introduced by Sparrow [6]. According to this criterion, two point sources are resolved if the saddle is flat and both picks are connected by a flat line in the intensity profile leading to K=0.473. Mathematically it is:

$$d = \frac{0.473\lambda}{NA}, \tag{5}$$

Houston criterion is another resolution criterion which is based on a definition that the resolution is the full width half maximum (FWHM) of the PSF. In this case the resolution is:

$$d = \text{FWHM} = \frac{0.515\lambda}{NA}, \tag{6}$$

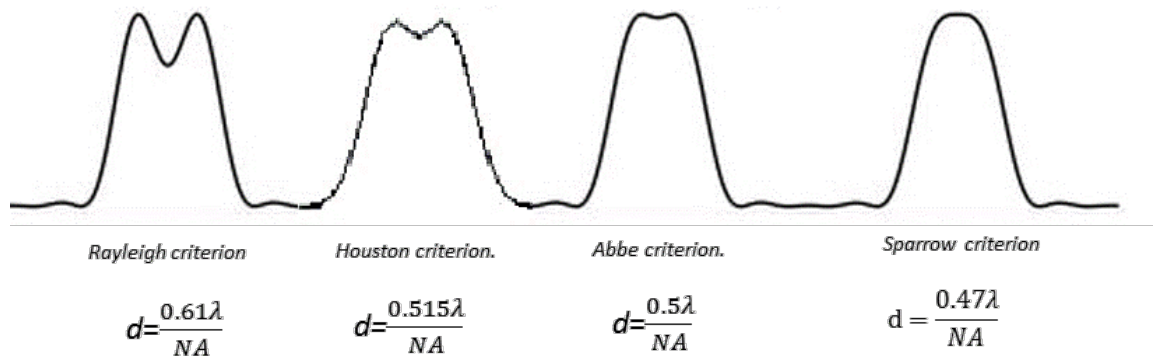

Figure 1.2: Different criteria [7].

By looking more accurate at the criteria definitions above one can notice the following points. First is that the Rayleigh's criterion is more suitable for the cases of coherent illumination when the image has a classical Airy disk shape with well pronounced zeros. In the case of incoherent illumination in most of practical cases the zeros of the intensity profiles are not pronounced that complicates the application of Rayleigh's criterion. In such cases, often Houston's criterion can be applied in practice since measurement of FWHM of the intensity profile can be usually performed in any cases including the incoherent illumination. Second is, all the criteria above are defined based on having two ideal identical point sources. Third is that in principle we can define resolution based on a single point source. In the latter case, the logical way of defining the resolution would be to use FWHM of the PSF divided by the magnification ($M$) of the optical system as a resolution of the system. It is easy to see that such definition corresponds to the Houston's criterion for the two-point object. In the following Sections, we will review some of the techniques developed to enhance the resolution of optical microscopes.

## 1.2    Microscopy and Nanoscopy Techniques

Abbe's work was critical and influential in this field [1, 2]. He introduced the fundamental formula for diffractive limit which has been the most important guideline for improvement of the resolution of microscopes. According to Eq. (1), to enhance the resolution one can shorten the wavelength of the illumination or increase the *NA* of the optical system. In 1904 Kohler constructed the ultraviolet (UV) microscope [8]. Later X-rays were used in biological microscopy and crystallography [9, 10]. Another achievement was electron microscopy where electrons with shorter de Broglie wavelength are used for microscopy [11]. *NA* enhancement was another way to decrease the diffraction limit and accidentally it began by the first immersion lens [12] even before introducing the Abbe's formula in order to decrease the aberration. All these developments led to invention of solid immersion lens (SIL) by the 1990s [13], since generally solid materials provide high refractive index even higher than liquids which are used in liquid immersion technique, see Figure 1.3 (a-c).

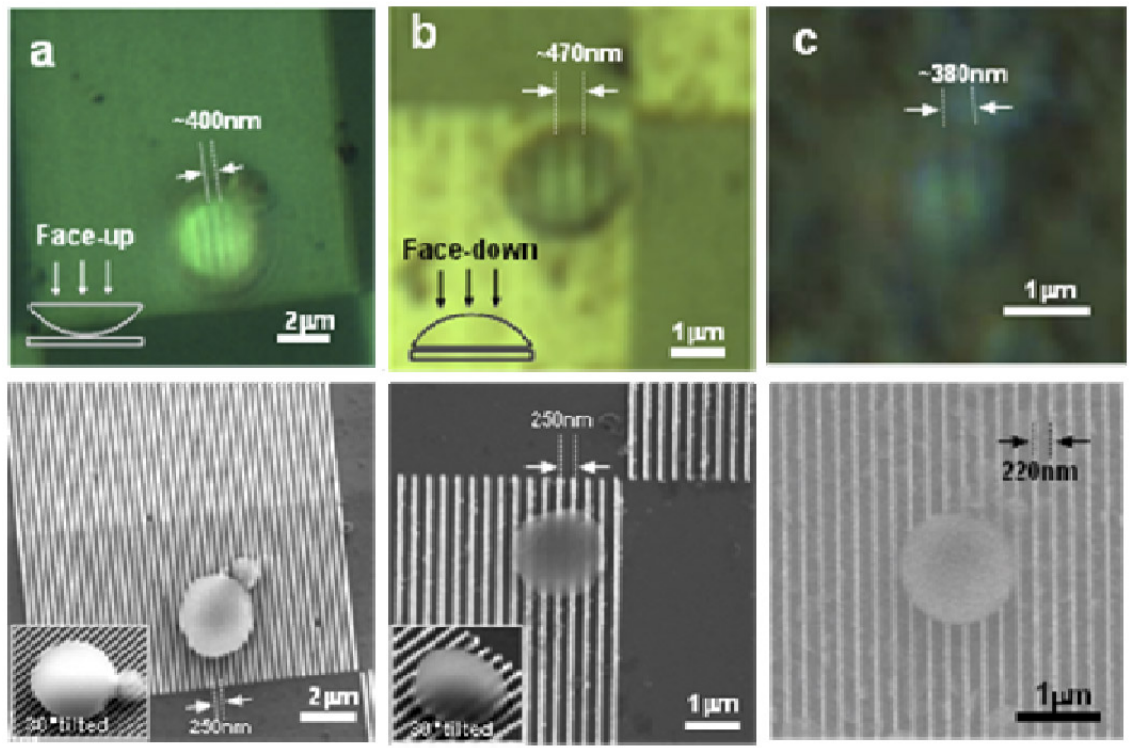

Figure 1.3: Optical microscope (top) with λ = 472nm, NA = 0.9) and corresponding SEM (bottom) images of individual nanoscale solid-immersion lenses (nSILs) sitting on metallic stripes separated by 250nm (a, b) and 220nm (c) [14]

Besides all the benefits of both strategies to improve the resolution by using Abbe's resolution formula, still there are some drawbacks:

- Using UV or X-ray has risk of damaging biological sample and also has very small penetration depth.

- Higher refractive index is always followed by higher absorption and chromatic aberration.

In 1950s by the advent of confocal microscope another step was taken towards better resolution by using Abbe's formula but still avoiding the drawbacks above. Using a laser and a pinhole into the optical path to block the out-of-focus light and make the PSF narrower [15], this technique can enhance both the lateral and axial resolutions approximately 1.4 times better than that of conventional microscopy, See Figure 1.4.

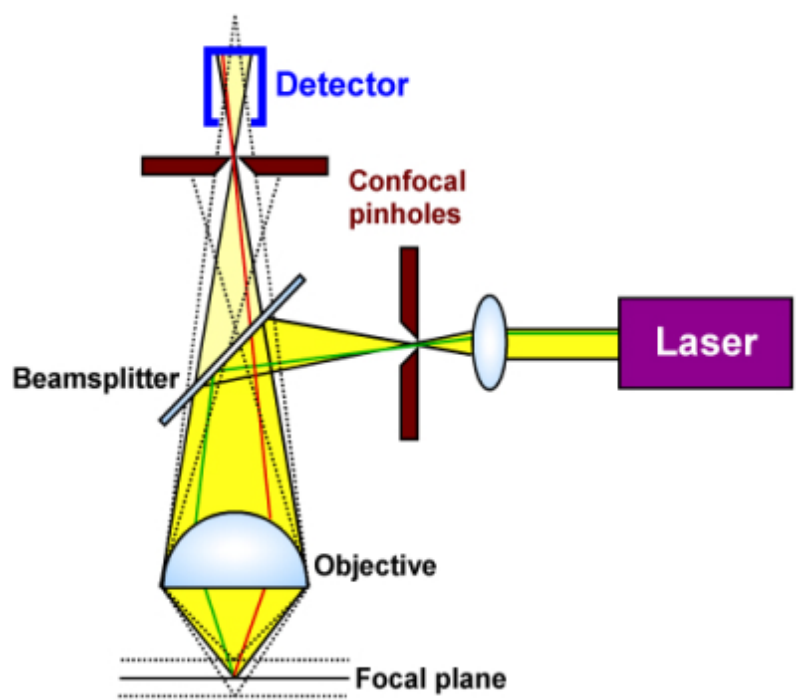

Figure 1.4: Typical design of a confocal microscope. It has two pinholes for illumination and detection of light paths. The second pinhole only passes light coming from focal plane of objectives to reduce the blurriness [95].

Although these techniques showed some resolution enhancement but diffraction limit was a bottleneck for enhancement of spatial resolution of microscopes which use visible light. Breaking diffraction limit in order to do imaging on sub-100-nm gradually became a hot topic. The term "super-resolution" has been used for the resolution of such techniques which can provide the resolution better than what the Abbe's diffraction limit suggests. Evanescent waves have been introduced as the key to break the diffraction limit since they have very short wavelengths compared to the wavelength of visible light which can interact with the object and provide the object's detail [2, 16]. In the next paragraphs we are going to briefly review some techniques which use evanescent waves.

The existence of evanescent waves was first theoretically postulated by Francia in 1942 [17] and the first experiment was performed in 1949. Despite propagation waves, evanescent wave is a near-field standing wave on a surface which its electric field decays exponentially with distance from the surface. The evanescent waves have wave vectors of the form [2]:

$$K_e = k_\parallel + i k_\perp; |K_e|^2 = \left|k_\parallel\right|^2 + |k_\perp|^2 = \left(\frac{2\pi}{\lambda}\right)^2 \tag{7}$$

where $K_e$ is the wave vector of the evanescent wave, $k_\perp$ and $k_\parallel$ are the wave vectors perpendicular and parallel to the surface respectively. The electromagnetic theory shows the parallel component of the wave vector of the evanescent wave is larger than the wave vector of correspondent propagation wave and can provide the finer detail, namely it can provide super-resolved image of the object.

In 1972 by the invention of near-field scanning microscope (NSOM) [18], the first super-resolution microscope based on evanescent waves came to exist. NSOM uses a probe to transfer light to the sample. The probe should be in the nanometric vicinity close to the sample. After scattering light the probe collects the evanescent waves which provide high spatial frequencies to resolve features below the diffraction limit, see Figure 1.5. This method needs scanning the area by the probe and it makes NSOM a complex and fairly slow microscope. This technique was an inspiration to develop other probe detection techniques such as scanning tunneling microscopy (STM) (1982) [19] and atomic force microscopy (AFM) (1986) [20].

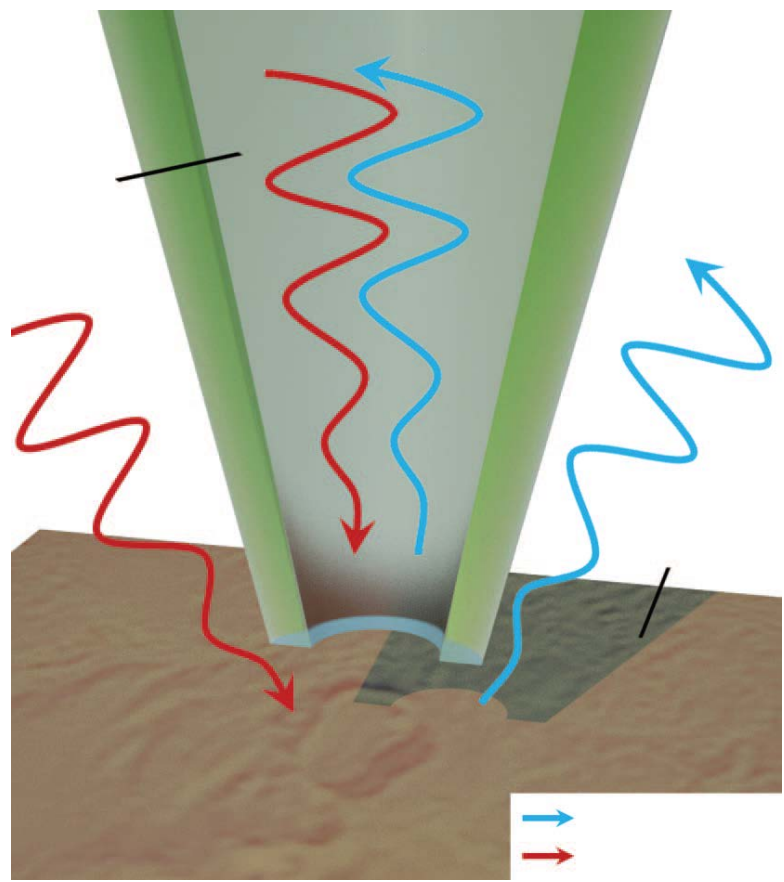

Figure 1.5: Schematic illustration of a probe and process of the near-field scanning optical microscope. Such probes can be used for both illumination the light and detection the near-field's evanescent waves [2].

Amplifying evanescent waves is another approach. Detecting evanescent waves before they decay totally, is another big challenge. In 2000, Pendry published a paper suggesting a slab of material with negative refractive index called "perfect lens" [21]. He showed theoretically that such a perfect lens is able to focus all Fourier components of a two-dimensional (2D) image. In other words, it can form the image using both propagation and evanescent waves. However practically there is a lack of natural negative index material in visible light. Some groups have tried to fabricate negative index refractive index material artificially. On the other hand some others have looked for alternatives. Pendry himself suggested that if we have a pure *p*-polarized light it can eliminate the dependency of the transition coefficient on μ, like negative index material which does the same by eliminating μ. In 2005, Zhang and his colleagues confirmed Pendry's conjecture using a silver superlens [22, 23], see Figure 1.6 (a). which could provide a subdiffraction-limited image on the other side of the lens. Zhang's group even went beyond that and two years later projected this image to the far-field and named the technique far-field superlens (FSL) [24], see Figure 1.6 (b). Xiong *et al.* developed the capability of FSL to 2D using a multilayer grating [25], see Figure 1.6 (c). Still the magnification of such subdiffraction image to the far-field was not possible until the advent of hyperlens [26], see Figure 1.6 (d). A sandwich-like half-cylindrical cavity (a piece of artificial meta-material) in the hyperlens will magnify the object while changing the evanescent waves into propagation waves. The wave vectors of the propagation waves gradually decrease while passing through the anisotropic meta-material so that subdiffraction information will be propagated and detected in air, see Figure. 1.6 (g).

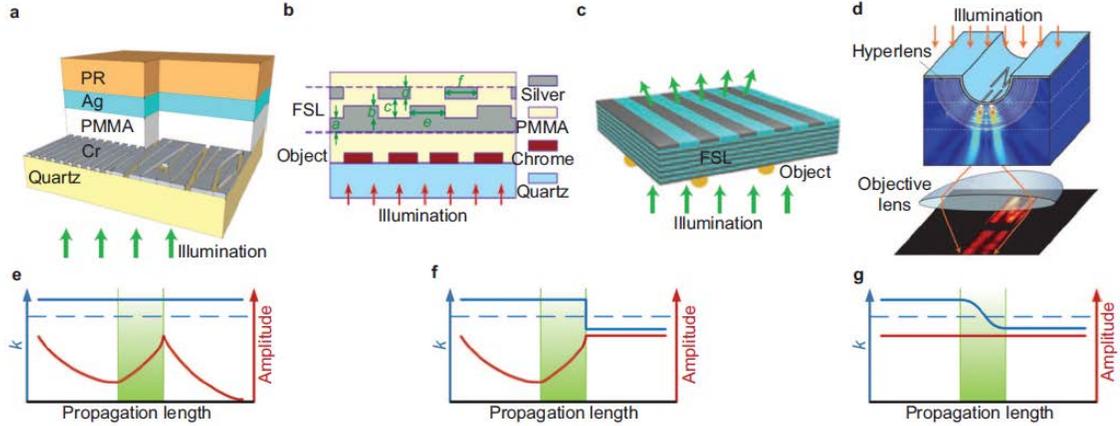

Figure 1.6: Schematic of (a) a superlens, (b) an FSL, (c) a 2D FSL, (d) a hyperlens, (e-g) Theoretical comparison of the lenses [22, 24-26].

Illuminating the sample with evanescent waves also can provide optical super-resolution. In this technique there is no need to amplify the evanescent wave like previous method. Illumination by the evanescent waves will be converted to propagation wave by being scattered from the sample and be easily detected from the far-field while it has the super-resolution information from the sample. For metallic samples surface plasmon polariton (SPPs) as the evanescent waves along the metal-dielectric interface play this role to extract the supper-resolution information due to the very short wavelengths they have compare to the incident light. The wave vector of SPPs is:

$$k_{sp} = k_0 \left(\frac{\varepsilon_d \varepsilon_m}{\varepsilon_d + \varepsilon_m}\right)^{1/2} \gg k_0 \tag{8}$$

where $k_0$ and $k_{sp}$ are the wave vector of excited light in vacuum and wavelength of SPP. $\varepsilon_d$ and $\varepsilon_m$ are dielectric functions of the dielectric and metal respectively. In 2005 Smolyaninov *et al.* demonstrated super-resolution imaging using SPPs [27, 28]. They etched the sample onto the gold film and deposited a microdroplet of glycerine onto the structure to create total internal reflection. It also acted as a magnifying mirror. They could see a magnified super-resolved image under the conventional optical microscope. SPPs only could provide super-resolved image if we have a metallic sample or a coated sample by metal. As another alternative to use evanescent wave, recently Hao *et al.* used near-field illumination by microfibers [29], see Figure 1.7 (a). They also could get super-resolution image of their sample without localized field enhancement. They showed for a sample with subdiffraction details illuminated by an external evanescent field, that somehow it was possible to shift the spatial frequencies to a propagation mode; consequently it was possible to projects the super-resolution information to the far field, see Figure 1.7 (b). They reported a resolution of 75nm approximately, see Figure 1.7 (c), although, for an arbitrary 2D pattern, this passive frequency conversion has not been able to reconstruct the image. To solve this problem, applying a series of recovery algorithms was needed beside the optical means.

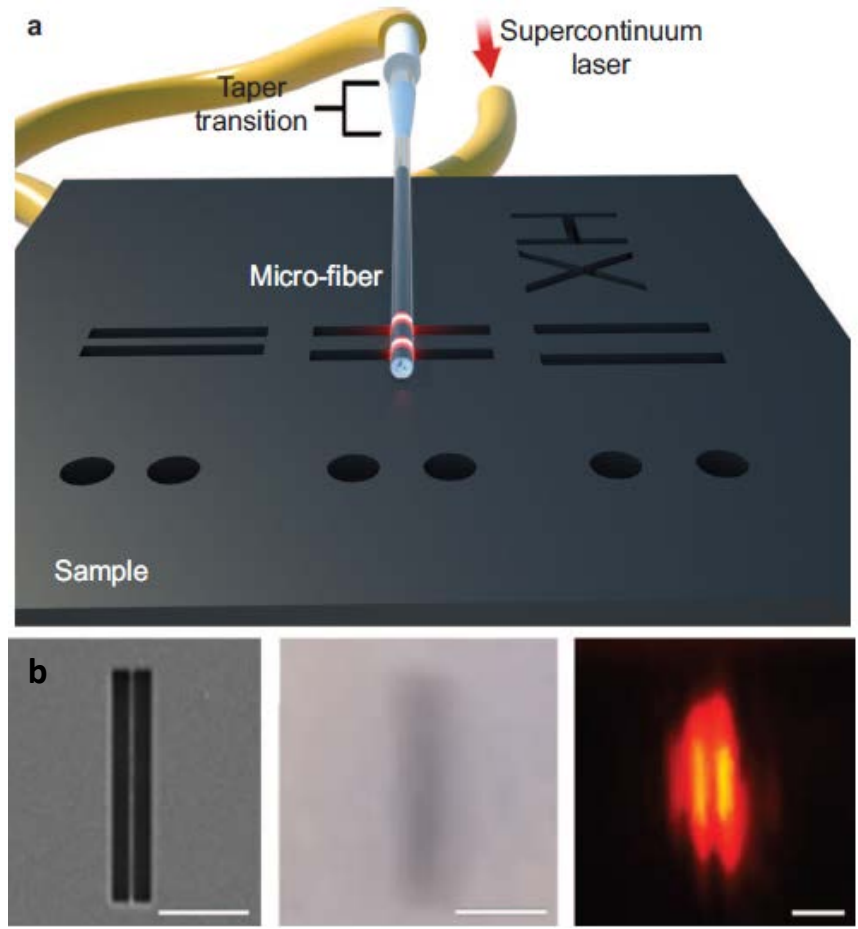

Figure 1.7: Microfiber-based nanoscopy. (a) Schematic view of microscope configuration. (b) SEM image of 75-nm-gap line pairs (left), wide-field optical microscopy (middle) and the image of microfiber nanoscopy (right). The length of the bar is 1 mm [29].

## 1.3    Microsphere-assisted Imaging

The new development in this area, which is closely related to the subject of this thesis, took place in 2011, when Wang, et. al, demonstrated a new nanoscopy technique based on using silica ($n$~1.46) microspheres 2<$D$<9 $\mu$m deposited at the top of investigated sample [30]. These microspheres act as microlenses which provide the virtual image of the sample captured by imaging through the sphere. For the purpose of showing the super-resolution capability of this technique, images of the stripes of a Blu-ray disk were demonstrated, where the stripes have a width of 200 nm and are separated by 100 nm, along with a nanometric scale star fabricated which had ~90 nm corners, see Figure 1.8 (a-c).

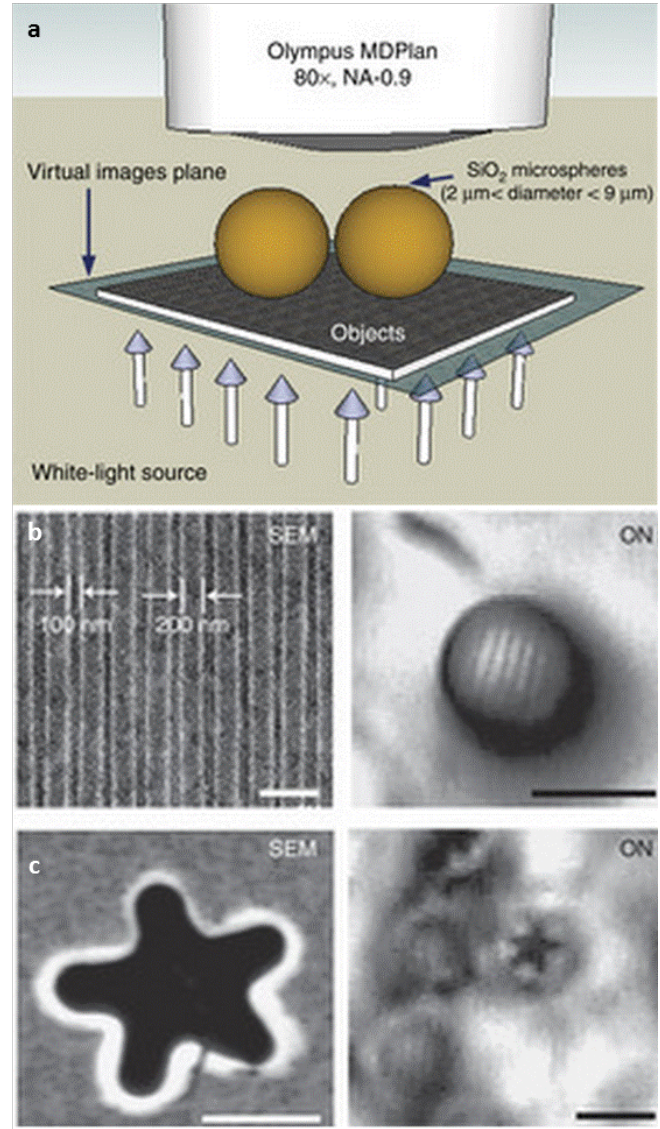

Figure 1.8: (a) Schematic image of white light microsphere-assisted microscope in transmission mode. (b) SEM image of a blu-ray disk shows the 200nm thickness and 100nm gap (left) and the super-resolved image by microsphere under the microscope. (c) SEM image of the star structure made on thin film for DVD disk with 90nm corners (left) and the super-resolved image of the star corners through a microsphere [30].

In order to explain the super-resolution capability of microspheres, they proposed that this phenomenon has the same origin which is extremely sharp focusing capability of small spheres termed "photonic nanojet" effect [31]. Due to reciprocity principle, sharper focusing is related to higher resolution imaging. It was also argued that this technique cannot be used in liquid and also for microsphere with $n$~1.8 or higher.

Lately, Dr. Astratov's Mesophotonics group has shed more light on this microsphere-assisted super-resolution technique [32]. They used fully immersed high-index barium titanate glass (BTG) microspheres with 1-50 µm diameter and index $n$~1.9-2.1 immersed in IPA with $n$~1.37. Using this technique, they demonstrated that this technique provides images with discerned minimal features on the order of $\lambda/7$ for relatively compact spheres with D<10 µm and for larger microspheres 50<$D$<220 µm the resolution drops to ~$\lambda/4$ ($D$ is the diameter of the sphere). The field of view of such images through microspheres linearly increases as a function of $D$. Lateral magnification also changes and has a peak for spheres with 5-10 µm diameters, see Figure 1.9 (e-d).

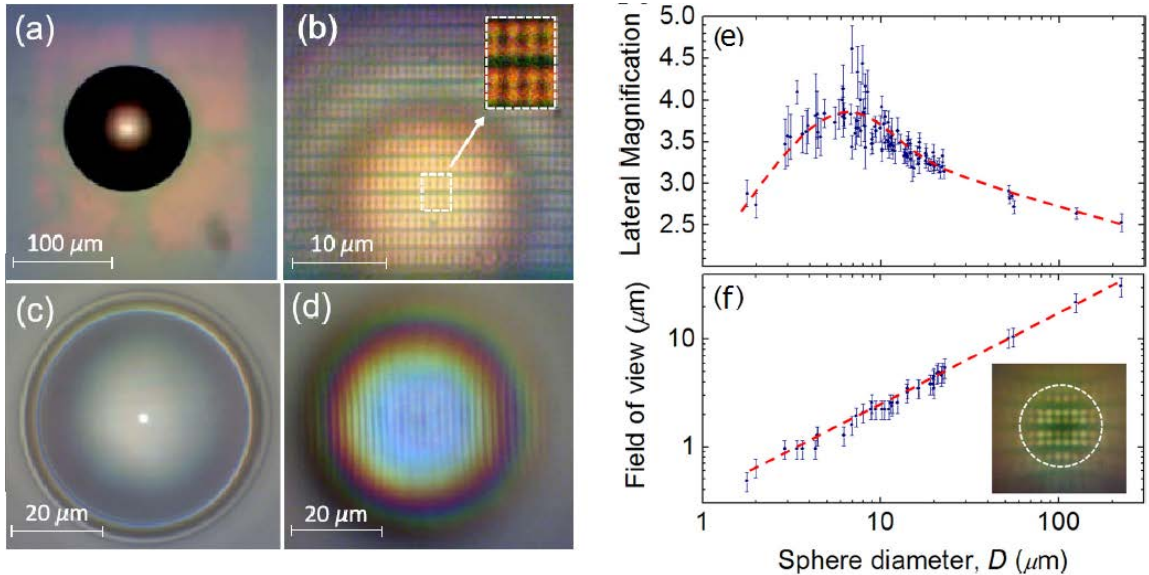

Figure 1.9: (a), (c) Two high index BTG microspheres fully immersed in IPA with the sizes of 100 µ and 20 µ. (b), (d) The super-resolved image of a nanoplasmonic array and Blu-ray. (e) Lateral magnification and (f) FOV obtained by BTG microspheres with $n\sim1.9$ as a function of $D$ [33].

A direct comparison of microsphere-assisted optical super-resolution microscopy technique with conventional optical microscopy and confocal microscopy were performed with microsphere size $D\sim15$ µm and $n\sim1.9$ and objective lens with $NA$ 0.95, see Figure 1.10 (a-d).

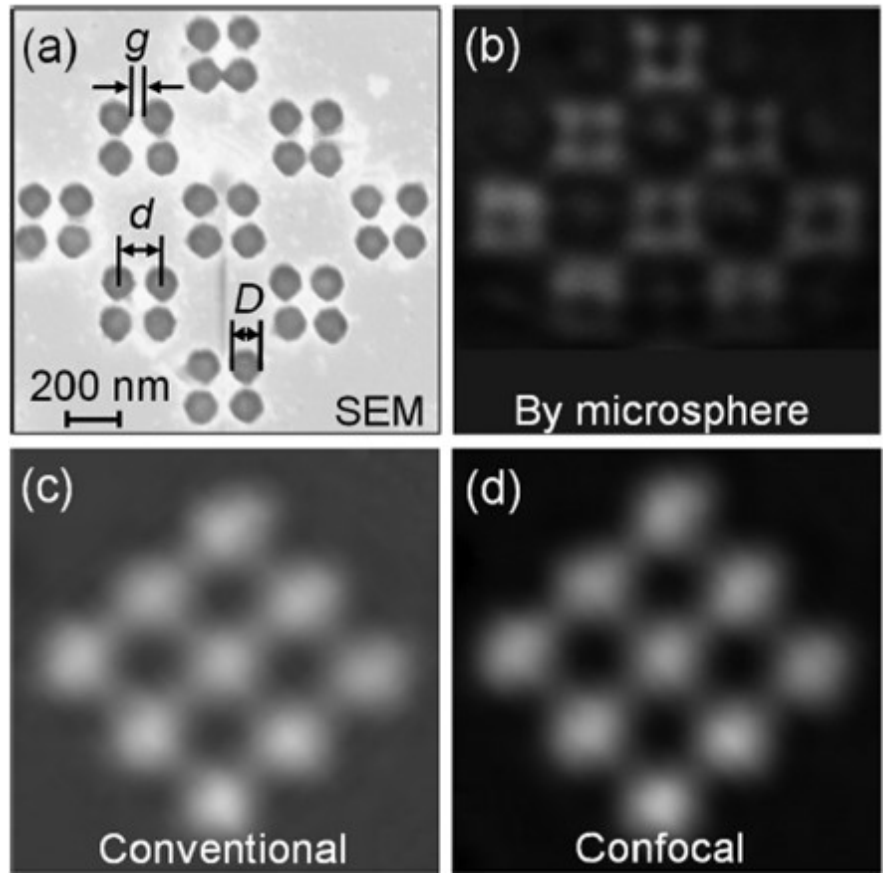

Figure 1.10: (a) SEM of the Au nanoplasmonic structure with gaps of ~50-60 nm and posts of 100 nm in diameter. (b) Super-resolved image of the structure by microsphere assisted technique. (c) Diffraction-limited image using conventional microscopy. (d) Confocal microscopy image of the same object. All of the images b-d were obtained through a 100x ($NA$=0.95) [34].

This comparison shows this technique provides superior quality and the higher resolution compared to both conventional and confocal microscopy. Another direct comparison with SIL technique also showed the microsphere-assisted microscopy provides superior resolution compared to millimeter-scale hemispherical SILs with the same index of refraction.

Investigating the role of the *NA* in microsphere-assisted technique revealed that drastic change in NA does not have much effect on the resolution of the image taken through the microsphere since the angle of light collected by microsphere is large enough regardless of the NA of microscope objective, whereas without microsphere NA of the objective plays an important role, see Figure 1.11 (a-f).

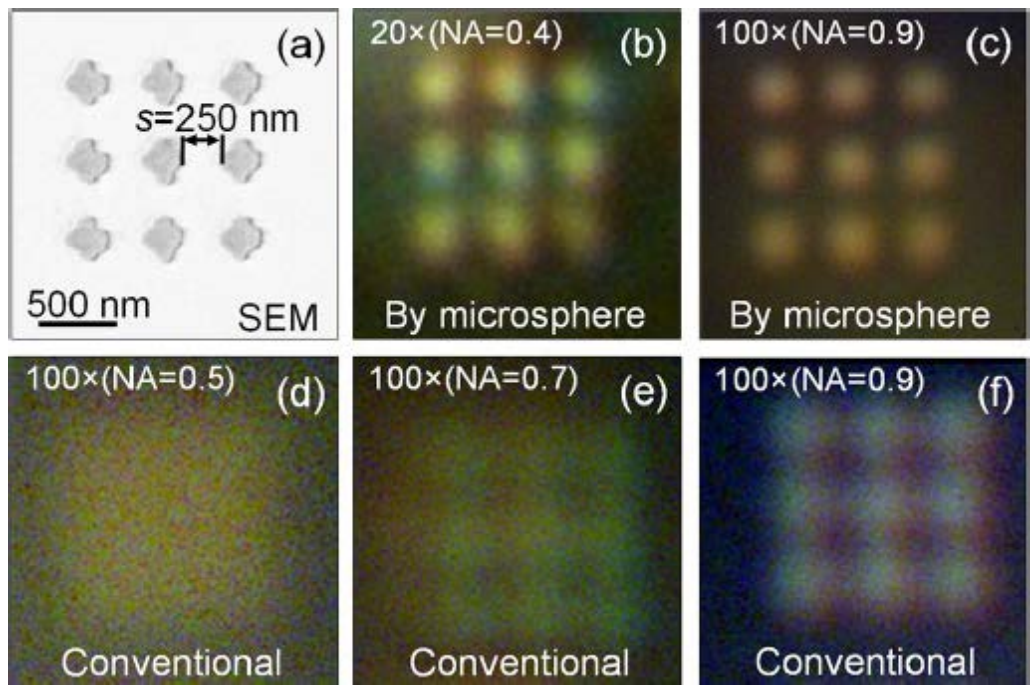

Figure 1.11: (a) SEM of the sample. (b), (c) Images through a 20 $\mu$m liquid immersed BTG microsphere (*n*=1.9) for NA=0.4 and NA=0.9 respectively. (d-f) Images from conventional optical microscope using different numerical apertures [34].

Beside some remarkable benefits of microsphere-assisted technique such as enhancement of magnification, resolution and easy to use, the broad application of this technique is under consideration which will be more discussed in Chapter 2.

As we already stated, a limiting factor for this method is represented by its small FOV. In order to expand the FOV of the image produced by these microspheres, in November 2013, Krivitsky et al. showed the locomotion of low-index microspheres provided by a glass micropipette [35]. Using micro-vacuum suction a silica sphere ($D$=6.1 μm, $n$~1.47) was attached to the tip of the micropipette. Their method made it possible to position the sphere with the accuracy of ~20 nm, See Figure 1.12.

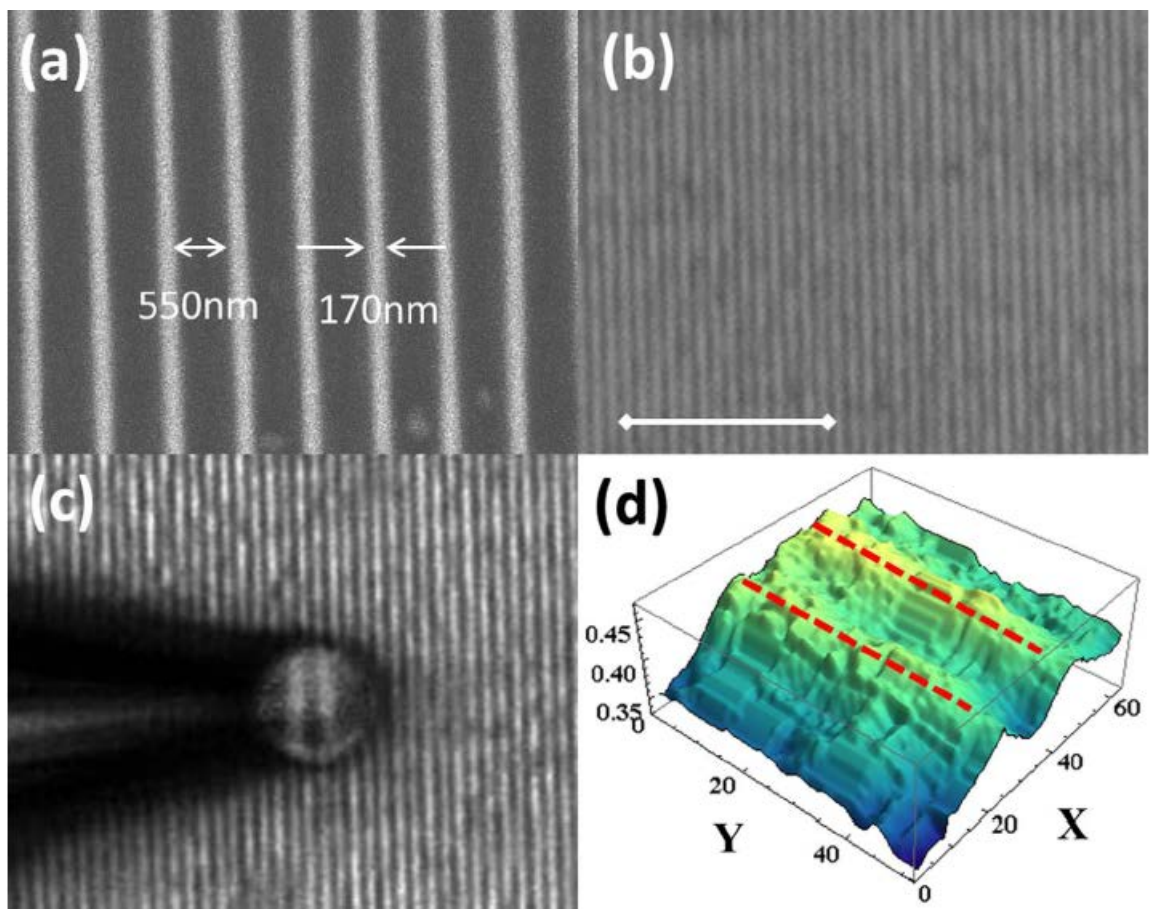

Figure 1.12: (a) SEM image of the sample. (b) Optical microscope image. The scale bar is 10μm. (c) Magnified virtual image using a microsphere attached to a micromanipulator. (d) Intensity profile of the magnified virtual image [35].

By the advent of        microsphere assisted technique, a competitive area in the optical super-resolution microscopy has begun. Several groups have attempted to investigate this technique from different aspects such as the physics of this super-resolution technique and resolution enhancement as well as applications. Confocal mode of imaging through the microspheres was demonstrated by Yan et al [36]. Basically, confocal laser scanning microscopy (CLSM) provides better resolution compared to conventional optical microscope [36, 37]. Claimed resolution generally have been based on "smallest discernible feature size" which can be considered as an intuitively understandable qualitative resolution criterion. However, its relevance to classical textbook resolution criteria needs to be established.

1.4     Resolution Measurement Procedure and Conclusions

In this section we review several methods for measuring the PSF of a confocal microscope as well as the basic theory of our method used in Chapter 3.

Two methods can be used to estimate the width of PSF of a CLSM [38, 39]. The first method is based on using standard slides which are specially designed for microscopy such as Richardson Test Slide Gen III and the MBL-NNF Test Slide [39]. The aforementioned slides have periodic line gratings of several spatial frequencies. There are two kinds of slides made for reflection and fluorescent mode separately. The resolution in this case is the spatial frequency at which the contrast of grating image disappears.

The second method is based on using very small beads of fluorescent polystyrene on the order of 200 nm, 100 nm or less as point sources with dyes suitable for measuring resolution at different wavelengths. The full width at half maximum of the pick (FWHM of the PSF) is the resolution of the system. Figure 1.13 shows a 3D image of a bead by a typical CLSM to measure both lateral and axial resolution of the system. Generally axial resolution is larger than the lateral resolution but in this work, we only discuss lateral resolution.

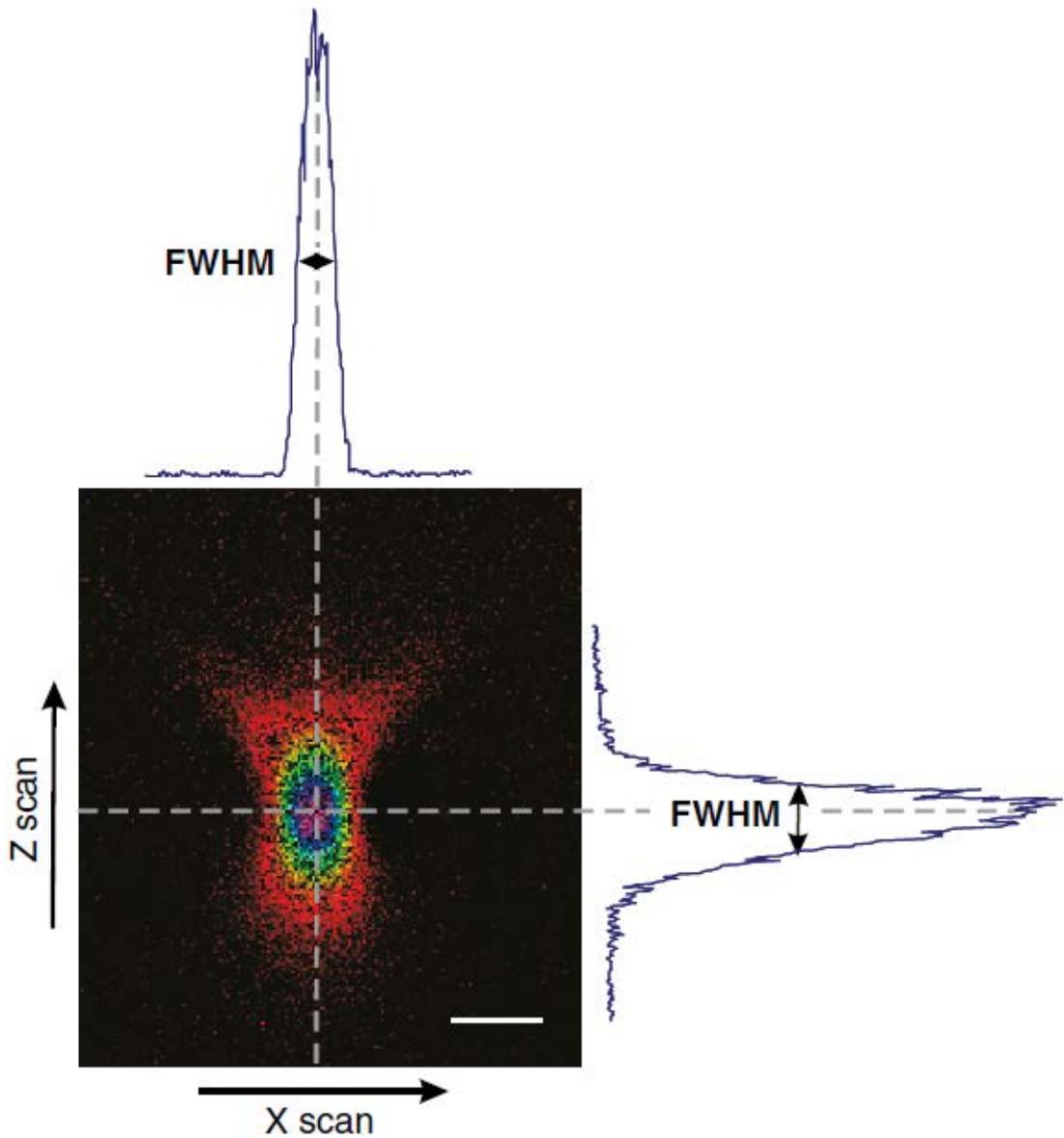

Figure 1.13: The image of a typical fluorescent bead from a fluorescent LCM shows both the lateral and axial PSF [39].

There is another method to measure resolution of the system based on analogy with the classical theory [40] where the image $I(x, y)$ is considered as a convolution of a diffraction-limited PSF and intensity distribution of the object $O$ $(u, v)$. The mathematical relation between $I$ and $O$ is:

$$I(x,y) = \iint\limits_{-\infty}^{\infty} O(u,v)PSF(u-x/M, v-y/M)dudv \qquad (9)$$

In which $u$, $v$ are the coordinates in the object plane and $x, y$ are also the coordinates in the image plane. The coordinates $x$ and $y$ are linearly related to the coordinates of object plane, while the coordinates $u$ and $v$ are generally defined in the object plane via magnification $M$ as $(u_0, v_0) = (x_0/M, y_0/M)$. The result of convolution, $I(x, y)$, is defined in the image plane and it is a computed image of the system. This result can be compared with experimental image. The width of PSF can be used as a fitting parameter in this procedure. Finding the best match between the computed and measured image would mean that we found FWHM of PSF. The resolution of the system can be obtained by dividing this parameter by $M$.

So far, it was shown during full liquid immersion, high-index spheres provide superior resolution on the order of $\sim\lambda/7$ in comparison to low-index spheres in air [37]. Initially, this resolution was established based on semi-quantitative criteria of discernibility of certain feature sizes in the optical images obtained through microspheres [32, 33]. After that, similar resolution of $\sim \lambda/6$ was estimated by using a simplified *one-dimensional* treatment with 100 nm rectangular functions separated by the gaps [34].Still, a compete treatment of the images in *two dimensions* which would be precise and applicable to objects with arbitrary shape still has not been developed in the previous work.

Also, using semi-quantitative criteria [29, 30, 32-34, 36]and in some cases using simplified *one-dimensional* methods of resolution estimation [34] it was demonstrated that microsphere-assisted technique's resolution exceeds that of SIL technology, conventional microscopies and confocal microscopies. Even with poor quality numerical aperture objective lenses (NA~0.4) this effect still exists. Beside all the benefits provided above by this technique, it still had serious limitations such as a small FOV and, poor control of positions of spheres. In 2013, a good step had been taken by Krivitsky et al. [35] with locomotion a microsphere, however the micromanipulator can damage the sample locomotion. On the other hand having access to a variety of microspheres of different sizes and having control over them by ease is still desirable.

To address these problems, we explore two approaches to developing microsphere-assisted imaging in this thesis. One approach is based on a technology development, so that we embedded high-index microspheres in a slab formed by transparent elastomeric material PDMS. This is described in Chapter 2 of this thesis. Such slabs attach easily to the sample and in a few seconds the microspheres come in the nanometric vicinity of the sample and provide super-resolution images. They also capable of be removed from the sample easily. We developed this technology in winter and spring of 2014 and published in a proceeding of conference NAECON in June of 2014. Another approach is based on developing resolution treatments which would be applicable to imaging of two-dimensional objects with arbitrary shape. This is described in Chapter 3.

Regarding standard methods of measuring resolution of optical system, one standard approach is based on using objects much smaller than the resolution of the system. Such objects create images which can be associated with PSF of the system. In this case, the resolution according to the Houston criterion, can be determined as FWHM of PSF divided by M. This simple approach can be applied to two-dimensional point objects (true "dots") as well as to one-dimensional point objects ("stripes"). The most important conditions for this type of resolution testing is have the object size well below the diffraction limit and to have sufficient brightness of the image for measuring it with good signal-to-noise ratio. These conditions are relatively easy to satisfy in diffraction-limited optics where semiconductor quantum dots, dye molecules and doped nanospheres, nanoplasmonic clusters, usually have dimensions well below 100 nm and sufficient brightness. In the case of optical super-resolution, however, it is much more difficult to find "good" point sources. As an example, for studying the resolutions about 50 nm only individual quantum objects such as single quantum dots or dye molecules would be sufficiently small to qualify as single point sources. However, brightness of such objects is usually prohibitively small for precise resolution measurements. Consequently, in order to circumvent these problems researchers usually use a different approach based on using relatively large-scale objects (sometimes millimeter-scale) with small size features (nanometer-scale) such as widths of metallic stripes or distances between the stripes or circles. In this approach only particular feature of the image such for example as the gaps between the stripes or between the circles can be sufficiently small. The resolution claims are made based on the observability or discernibility of this feature.

It should be noted that this is much more subjective technique requiring more detailed analysis. In what follow we show that this approach can lead to greatly exaggerated resolution values. Another problem is connected with the gravity and Brownian motion which can cause the beads move out of the image area which is limited by a circle with *D/4* diameter centered with the vertical axis of the microsphere or push it to the edge of this area, which means having a bad result with low quality image because of the spherical aberration or pincushion effect.

In Chapter 3, we presented the classical theory of image creation formulated in Eq. (9), however we allowed subdiffraction-limited FWHM of PSF. Conventionally, Eq. (9) was used only in the domain of diffraction-limited optics. We are not aware about previous use of this approach for the resolution treatment of super-resolved images. In this work, however, we suggest to extend this approach in the domain of optical super-resolution by allowing the PSF to take subdiffraction-limited widths. Thus, our approach does not allow to identify the mechanism of super-resolution. However, it allows determining the super-resolution in a way consistent the classical text-book definitions well accepted in diffraction-limited optics. In some sense, our resolution treatment represents a phenomenological approach to this problem. We demonstrated this approach to measure the super-resolution of an experimental image from a microspherical lens. We also developed this method to measure the resolution of cylindrical lens.

CHAPTER 2: FABRICATION OF PDMS THIN FILM WITH EMBEDDED
MICROSPHERES

2.1    Introduction

Since the adventure of microsphere-assisted microscopy, some groups have tried to develop this technique [2, 30, 32-36, 41, 42]. Although this is a very simple way to obtain the super-resolved images, the broad applications of this method are somewhat complicated due to several factors. First, the super resolved image obtained through microsphere are limited to FOV ~$D$/4 which $D$ is the diameter of the microsphere [33].

Second, in order to get super-resolved image, the smaller microspheres about 10 µm

diameter or less are needed [37]. The reason why the super-resolution can be attained

with relatively small, almost wavelength-scale spheres is still debated in the literature

[43, 44]. Actually, the smaller sphere is, closer contact with nanoscale object can be

achieved. However, there might be other more fundamental factors responsible for this behavior such as the fact that sharper focusing can be performed only with smaller spheres [45]. For now, the need of use relatively small microspheres still remains largely an experimental fact. However, in any case, it would mean that field-of-view (FOV) is rather limited in this technique. Third, the microsphere immersed in liquid generally is not stable and moves with micro-fluidic flow; therefore there is no control over it. Forth, after evaporation of the liquid, there are some microspheres which stuck to the sample.

The mechanism of such stickiness is not completely clear, and it can be different for different spheres, including charging effect, Van der Waals forces, and chemical interaction with the substrate. In order to tackle these drawbacks, Krivitsky et al. have used a micromanipulator with an attached microsphere in order to probe the desired part of the sample [35]; although it provides some control, but, the sample can be easily damaged by micromanipulator during landing and locomotion. In addition, such microprobes are usually fragile and precise positioning of microsphere usually requires complicated additional 3-D micromanipulation kit. On the other hand, handling individual microspheres is not an easy task. In some cases, optical tweezers can be used [46] to position microspheres. However, in any case, such manipulation requires additional equipment which somewhat detracts from the simplicity of this method.

In this work, as an alternative solution of this problem we fabricated microsphere-embedded PDMS thin film with almost the same refractive index contrast as silica microspheres in air in order to obtain super-resolved image while avoiding the above listed difficulties. Our approach is based on an idea that such thin films with embedded microspheres can be translated along the surface of the sample to align different spheres with objects of studies. However, our approach involved several hurdles which we had to overcome to achieve good results.

To this end, PDMS is a very well-known material which satisfies the desired optical and mechanical properties which are interested [47-49]. Some of those properties are as follows:

- Transparent for visible light with n=1.4.

- Highly flexible.

- Moderately viscoelastic.

- Durable in flexibility and viscoelasticity for several uses.

- Chemically inactive.

Historically, microspheres were embedded in PDMS films by Whitesides group [47] and by the Mesophotonics group (Prof. Astratov's group) at UNCC [48]. There are mainly two features of this technology which are special for this work and make it different from the previously published results. First is that in order to realize right conditions for virtual imaging we needed to provide a certain refractive index contrast between the PDMS and embedded spheres. This index contrast is about 1.4-1.6 range. Taking into account that the index of PDMS is close to 1.4, it means that we had to use high-index spheres with n~2.0. Second, is that for the purpose of imaging we had to assemble microspheres very close to the lower surface of the PDMS films. This is required for picking object optical near-fields by the microsphere. For these reasons, we are interested in fabricating a PDMS (Sylgard 184, Dow Corning) thin film with embedded high index microspheres touching the thin film surface or protruding from it.

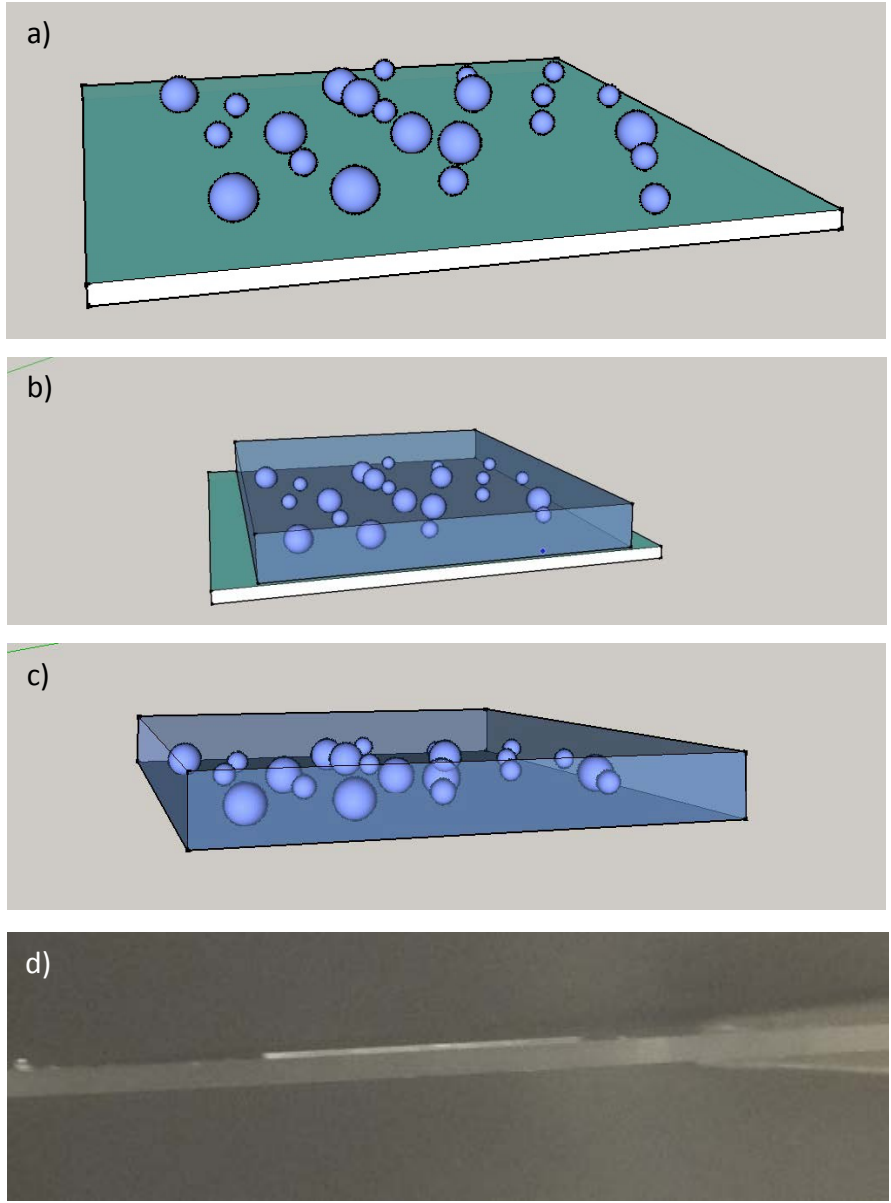

Figure 2.1: Step-by-step fabrication of PDMS thin film with embedded spheres. (a) Deposition of microspheres on glass substrate. (b) Casting and curing PDMS layer. (c) Lifting the layer off.

## 2.2    Fabrication

The fabrication is a 3-step process illustrated schematically in Figure 2.1 (a-c).

Figure 2.1 (a) shows some high index microspheres (n~1.9-2.1) with different sizes are deposited on a cleaned glass substrate. Then the PDMS is provided with the base/curing agent ratio of 10/1 successively and well mixing process. The mixing of PDMS leads to creation of microbubbles throughout the volume. Leaving the PDMS for enough time let the bubbles escape from the surface of the PDMS. Then the PDMS is casted over the spheres, see Figure 2.1 (b). After that, it is heated in 200F in a mini oven for about 1 hour. In order to take the PDMS off the glass substrate, a thermal treatment is used by putting the samples in a freezer or using ice. Figure 2.1 (c) shows schematically the PDMS with embedded microspheres taken off from the surface by a few minutes of thermal treatment.

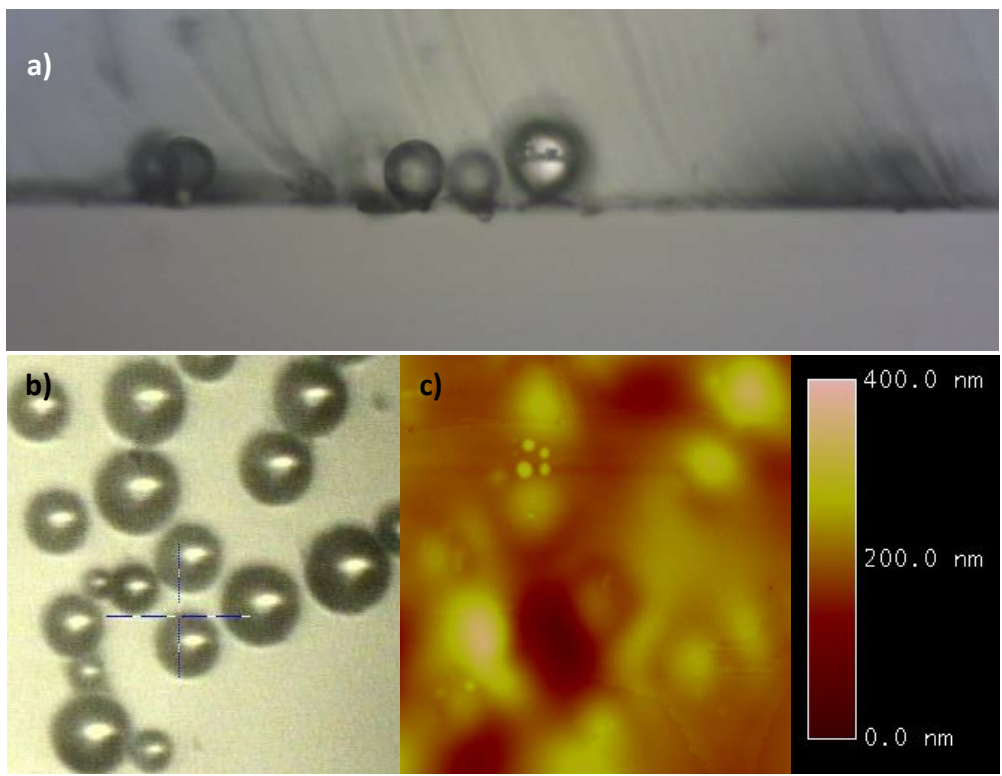

Figure 2.2: Physical test of fabricated PDMS layer is done under conventional and AFM microscopes. (a) Side view of the PDMS layer. (b) Top view of the layer. (c) AFM of the same spot in (b).

## 2.3   Characterization of the Fabricated Structures

Figure 2.2 (a) is shown a side view of a fabricated PDMS layer cut by a scalpel and viewed under a conventional optical microscope. It shows all microspheres are embedded at the bottom side of the layer. The edge of the PDMS slab looks rough in this image; however it reflects the quality of cutting by the scalpel. The actual quality of surface of PDMS slab is much higher.   Figure 2.2 (b) is the top view of such microspheres embedded in the PDMS layer. Figure 2.2 (c) is the AFM image of the same spot in Figure 2.2 (b). Characterization by AFM was done in collaboration with Kenneth W. Allen. It shows that the surface bumps coincide with the positions of the spheres that basically indicates that the spheres are located in nanometric vicinity of the surface. In some cases the microspheres were found to protrude through the PDMS slab. Generally, estimating the gap separating the spheres from the surface is a complicated task and we did not have sufficiently precise methods for this type of characterization. One approach of estimating the gap is to measure the magnification. The derivation of magnification formula below was done separately by the author and Kenneth W. Allen, This derivation is based on geometrical optics analysis:

$$M(n', r, g) = \frac{n'}{2(n'-1)\left(\frac{g}{r}+1\right) - n'} \qquad (10)$$

where r is the radius of the sphere, $g$ is the gap between sphere and the object and $n' = (n_{sphere}/n_{medium})$. Eq. (10) was obtained in the limit of geometrical optics. This means that it can only be used with sufficiently large spheres such as $D$>10 µm. Generally, these studies showed that the spheres are separated from the surface by gaps which are smaller than 100 nm. However, the accuracy of determination of parameter $g$ using Eq. (10) drops for smaller gaps.

Studies of AFM bump maps, also showed that on average the spheres are extremely close to the lower surface of the slab with the separation being much smaller than the wavelength of light. Therefore this test shows the PDMS layer is ready to use and a good number of microspheres which protrude the PDMS can make good contacts with the sample in order to provide the best quality image.

# CHAPTER 3: METHOD OF RESOLUTION TREATMENT

## 3.1 Introduction

Following the fabrication of microsphere-embedded PDMS thin film, as we discussed in Chapter 1, in order to report an accurate and reliable resolution of our microsphere-assisted microscope, we need to use a method based on textbook definition of resolution. A conventional way of defining the resolution would be based on using "point" sources. The width of the image of point source divided by $M$ can be assumed as being a resolution of the system. In practice, this approach can be applied to various nanoscale objects such as quantum dots or dye molecules. The imaging of such objects, however, can be difficult because of their vanishingly small intensity. Alternative way would be to use different large-scale objects which have some small-scale details with a recognizable shape. Sometimes the researches tend to conclude that if a given feature is seen by the imaging device, this feature "is resolved" which means that the resolution of the system is equal to the size of this feature. Although this is intuitively understandable logic from the first sight, it requires more rigorous mathematical justification. It can only be done based on the textbook first principles. Developing such mathematical apparatus allowing treatment different images and aimed at defining their resolution was a main task of this Chapter. It also provides a reliable way to compare different results reported by other groups that are reported based on observability of "minimal discernible feature sizes". We show that the latter criterion is not reliable and in some cases can be misleading. In this chapter we are going to introduce our method. This work was mostly begun by collaboration with Yangcheng Li and Kenneth W. Allen at the very beginning. The development of this method by doing several simulations and the analyses was done by the author in this chapter.

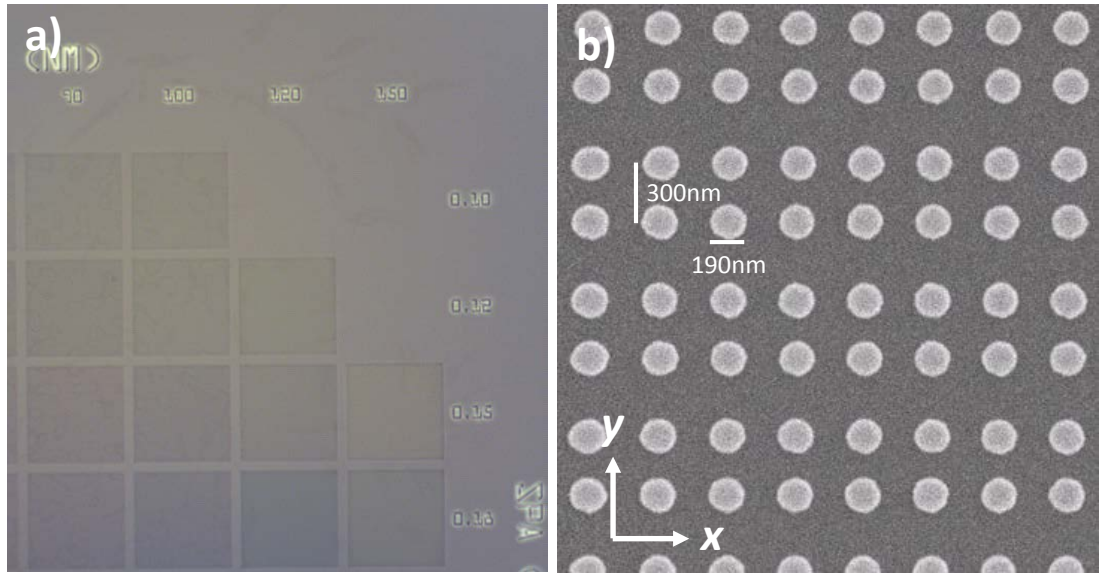

Figure 3.1: (a) Optical microscope image showing the large area of the sample with multiple arrays of the gold nanostructures. Each square represents an array with different dimer size and separation. (b) SEM showing a typical array of dimers.

## 3.2 Nanoplasmonic samples

In order to investigate super-resolution of our system we need a well characterized nanostructure sample as a reference and known object. It should contain small structures with typical dimensions on the scale of a diffraction limit or smaller. We have used gold and aluminum nanostructure samples which were fabricated at Air Force Research Laboratory (AFRL). These samples contain several arrays of dimers. Each dimer consists of two closely spaced metallic cylinders which are seen as circles in Figure 3.1 (b). Figure 3.1 (a) shows a typical overview of such a sample. Each array contains a large number of dimers with different diameters and separations. They were assembled in 2D with periods of 700 nm and 350 nm respectively in *x* and *y* directions. Figure 3.1 (b) shows a typical array of dimer with the dimer diameter of 190 nm and 300 nm center to center separations. The heights of the metallic cylinders (along *z*-axis) are about 45 nm. The fabrication is performed on a sapphire substrate, by the lift-off process technique. First, a 10 nm titanium layer is deposited on the sample. Second, a sacrificial resist is deposited on the Titanium layer. Third, an inverse pattern is etched on the resist by electron beam lithography. Forth, gold or aluminum will be deposited all over the sample. And at last, the resist with the metal deposited on it is developed ("is washed out") using a solvent. So the remaining pattern is the metallic structure on the titanium layer.

### 3.3    Resolution Treatment Method for 2D Lens (Spherical Lens)

As we discussed in Chapter 1, the image of an ideal point source object is defined as PSF. Different criteria (Abbe, Rayleigh, Houston and Sparrow) were discussed which are all sufficiently close to ~$\frac{0.5\lambda}{NA}$. Among these criteria Houston criterion has been used for measurement of CLSM resolution. There might be a very good reason for that. The prevalent method discussed in Chapter 1 which uses beads as the available point sources, make it possible to obtain the PSF directly. And Houston criterion is the only criterion which has a simple direct relation to PSF, and it is the width of the PSF divided by *M* which is the resolution itself. Thus, we decided to follow the Houston criterion in our work. It should be noted, however, that the point sources are difficult to find in super-resolution regime (see Chapter 1). Many objects such as Blu-ray disk, dye-doped microspheres, nanoplasmonic dimers and bow-ties with typical dimensions on the order of ~100 nm would not qualify as "point sources" for systems providing resolution, for example, about 50 nm. If the object is not point-source, then it was a procedure in diffraction optics for treating such images based on convolution of the object with PSF.

To investigate the resolution of our microsphere-assisted microscope, we have used the classical theory formula Eq.9:

$$I(x,y) = \iint\limits_{-\infty}^{\infty} O(u,v) PSF(u - x/M, v - y/M) du dv \qquad (11)$$

Figure 3.2 (a) shows the schematic diagram of an actual image formation process through the microscope step-by-step with a typical dimer sample as the object. It is very important to remind that we are not allowed to apply any of the criteria formulated for two point-objects, since the dimer contains two circles with a *finite size*. Figure 3.2 (b) shows the image reconstruction and resolution treatment step-by-step based on the analogy with the established image formation process according to the classical theory. Having SEM image of the nanostructure sample, we have drawn the idealized object. Our approach to the image reconstruction is similar to classical diffraction-limited optics in a sense that we convolved the drawn image with Gaussian PSFs. The convolution was repeated many times using PSFs with different widths. In order to find the most similar image to the experimental image we have checked the similarity of intensity profile of each calculated image to the experimental image. We have used Houston criterion for defining the resolution. According to this criterion, FWHM of the PSF which provides the most similar intensity profile divided by $M$ is the resolution of the system.

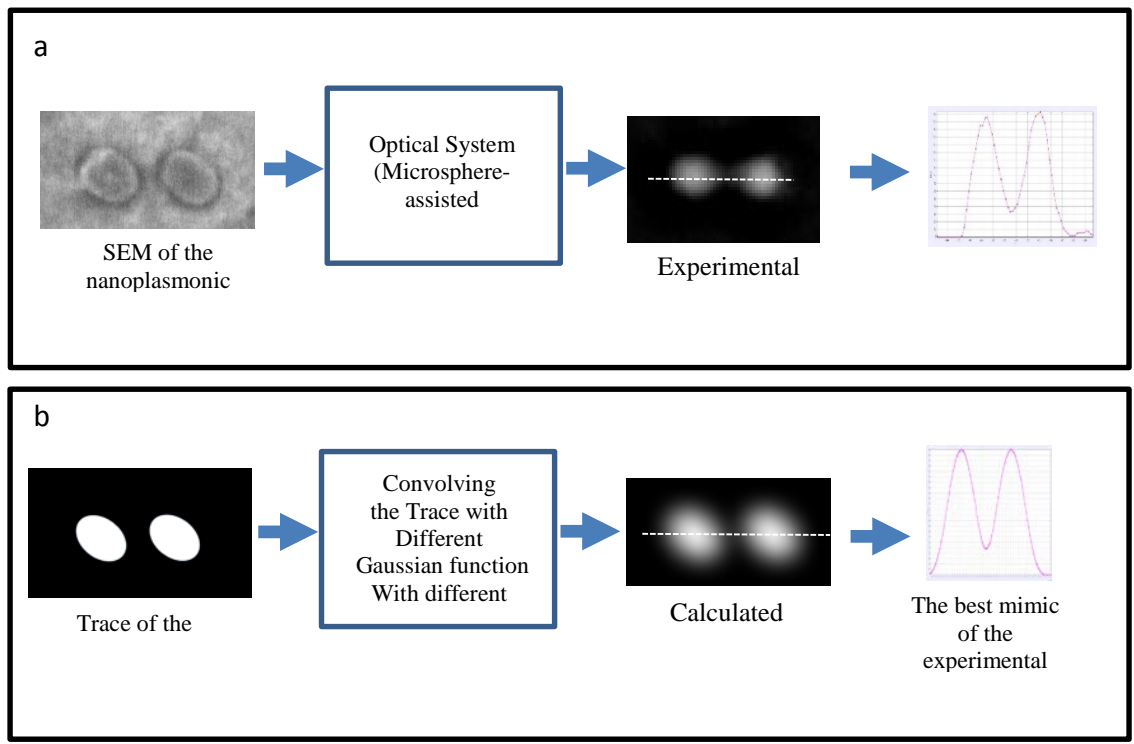

Figure 3.2: Comparison of image creation processes through a typical microscope and through the model. (a) Shows the step-by-step process of image creation through a microscope. (b) Shows the step-by-step process of image creation through the suggested model.

## 3.4    Effect of Nanostructure Size on the Precision of the Method

During fitting process, one important factor which needs to be taken into account is the diameters of nanocylinders and their separations in the dimers which we selected for investigating the resolution of our system. In this Section, it is shown how this would affect the precision of determination of resolution by this method. It should be noted that the basic principle of convolution with PSF does not depend on the characteristic dimensions of the object. However, the "sensitivity" of the image to the changes of the width of PSF is dependent on the size of the object. The basic idea is that if the object features are much larger than PSF width then the image should not be sensitive to the variations of PSF. In the opposite limit we might expect that the calculated image would be looking similar to PSF. However, if the object is too small, it becomes too week and difficult to observe. So, the task was to determine the range of variation of object dimensions which would produce most visible changes in the quality of the image and for this reason would be most suitable for undertaking an experimental study of the resolution of the system. The difference with conventional diffraction-limited optics, though, was that we generally are not restricted to the PSF widths determined by the fundamental diffraction limit.

To show how the size and separation of the nanostructure would affect the precision of the final calculated resolution, we selected two cases for this purpose. We consider an idealized drawn dimer with 240 pixels diameter size and 390 pixels center to center separation. We used two small PSFs compare to the size of the dimer. Figure 3.3 (a) shows the intensity profiles of the convolved images for two different PSFs with the FWHMs 30 (black) and 60 pixels (green). Since it is a hypothetical example, there is no real profile to represent.

By looking at these profiles one can see they have similar shape, but slightly different steep around the flat top and the flat down areas. It shows that neither FWHM nor the slopes of the picks are sensitive enough to change of the PSF; therefore, these calculations support our initial assumption that the large-scale objects are not suitable for defining the resolution of our system experimentally.

In contrast, Figure 3.3 (b) shows the case for PSF with the FWHM about the size of the dimer. We used the same idealized drawn image which used in Figure 3.3 (b). For the black curve, the PSF width is equal to 120 pixels and for the green curve, the PSF width is equal to 240 pixels. It is seen that for a green curve, the 240 pixel PSF width is comparable to the 240 pixel diameter and 390 pixel separation of the dimer. Whereas for a black curve, the PSF width is 120 pixel which is two times smaller than these characteristic object dimensions. As we can see, there is a change in FWHM of the picks but there is also very noticeable change in the saddle-to-pick (S/P) ratio for the intensity profile.

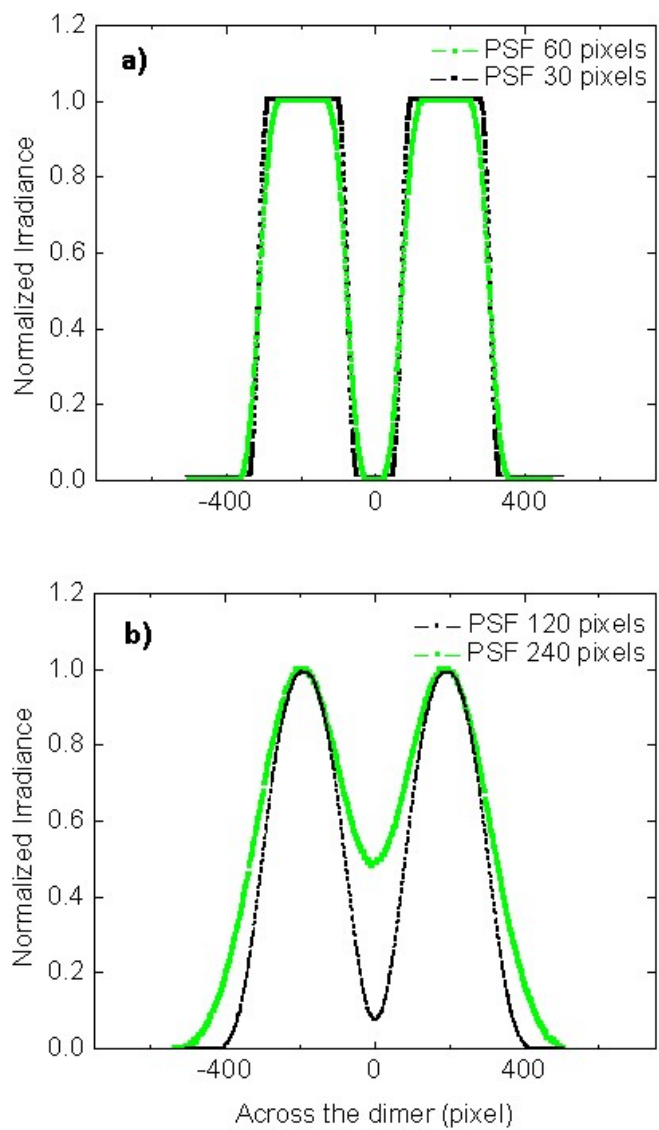

Figure 3.3: Comparison of two calculated intensity profile for (a) 390 pixel dimer diameter size. Green represents PSF with 60 pixel width and black represents PSF with 30 pixel width. (b) 390 pixel dimer diameter size. Green represents PSF with 240 pixel width and black represents PSF with 120 pixel width.

Considering two different cases in Figure 3.3 (a), (b), one can notice 4 important outcomes. First, the size of the nanostructure has significant effect on how strongly the calculated image depends on the PSF width. Second, the most markedly pronounced variations of the image can be detected for the PSF widths comparable with the typical dimensions of the object. Third, S/P ratio and FWHM of the profile respectively are good parameter to compare experimental and calculated profile. Fourth, further reduction of the size of the object leads to the image quality deterioration. It can be explained by the reduced intensity of the image. In principle, the image can still be calculated by increasing as an example the time of computations. However, experimentally the image may become two weak at some point. Due to a finite signal-to-noise ratio any imaging system would become incapable for recording such weak images of extremely small nanoscale objects. On the other hand, the image of two extremely small size objects calculated in this limit will be similar to superposition of two PSFs. So, the calculated image would be sensitive to variation of PSF width, but insensitive to variation of the size of extremely small objects in this limit. To summarize, the best tradeoff for experimental studies of resolution seem to take place in the intermediate regime when the objects has dimensions comparable with PSF, on the other hand we do not have the PSF of the system. In order to solve this cyclic problem we need to do some modelling.

3.5    Analysis to Find the Optimum Nanostructure Size

Since we have variety of dimer sizes and separations, we need to know which array is the best for measuring the resolution. Our goal is to find an array which has both sensitive S/P to change of PSF and has a wide range of S/P from zero to one for the PSF range of 25nm to 200nm, since 25 nm is the best resolution which can in principle be expected based on published work [36] and 200 is ~$\lambda$/2 diffraction limit of the LEXT OLS4000 confocal microscope with 405nm laser available in the cleanroom of our department.

Our samples, have arrays of dimers with diameter range 90 nm to 190 nm and separation range 150nm to 300nm. So, the minimal dimensions were limited by 90 nm and 150 nm for the diameter and the gap, respectively. For our analysis, however, we studied three cases with the different diameter sizes 50, 100 and 200 pixels, and with different edge to edge gaps of 25, 50 and 100 pixels. Different combinations of these parameters form 9 different structures. We have examined each structure by 4 different PSFs 25, 50, 100 and 200nm. As we discussed earlier, calculations with different PSFs result in differently looking images. In order to compare these images, we paid attention to their overall appearance (flat top or round top, S/P ratio and total width of double-peak structure), however for quantitative comparison we focused on the S/P ratio which generally represent the "quality" of resolution of these two circular objects with finite dimensions. Figure 3.4 (a) shows the S/P versus dimer diameter for the case of 25 pixel edge-to-edge gap. Such a graph shows both sensitivity of S/P to PSF and the range of PSFs which provide 0<S/P<1. For instance, for the dimer with 50 pixel diameter and 25 pixel gap, the range of S/P begins with 0.2 for 25 pixel PSF and ends for 100 pixel PSF. Figures 3.4 (b), (c) respectively show the results for 50 pixel gap and 100 pixel gap for the same dimer sizes.

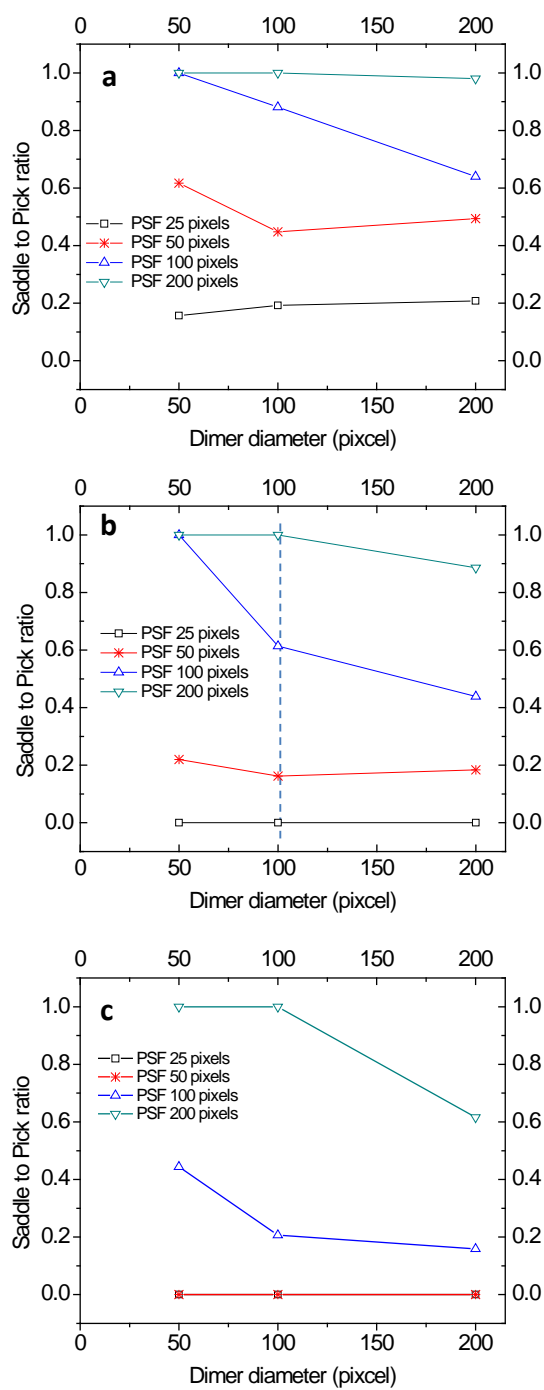

Figure 3.4: Graphs show S/P versus dimer diameter for different PSFs. (a) for dimers with 25 pixels edge-to-edge gap. (b) for dimers with 50 pixels edge-to-edge gap. (c) for dimers with 100 pixels edge-to-edge gap.

By analyzing the S/P ratio dependencies presented in Figure 3.4 (a-c), one can conclude that for the array containing dimer with 100 pixel diameter and 50 pixel gap S/P ratio changes from 0 to 1 for PSF increasing from 25 pixels to 200 pixels. It also shows that the sensitivity of S/P ratio to the width of PSF is not linear. Generally, these graphs confirm our initial assumption that the maximal "sensitivity" of the S/P ratio indeed occurs for the objects with the characteristic dimensions comparable with the PSF width. In order to select the nanostructure for the resolution studies we first estimate the diffraction-limited resolution of our system based on the Abbe formula which gives us something on the order of $\lambda/2 \sim 200$ nm. After that, we make an assumption that the super-resolution (resolution better than diffracton limit) is possible and estimate the expected range of resolutions as $\lambda/4 \sim 100$ nm. The super-resolution would correspond to sub-100 nm values. According to our analysis this means that the array of dimers with sub-100 nm dimensions would be the best. In practice, however, it simply means that we should select the array with the smallest dimers with 100 nm diameter size and 50 nm gap; therefore this array will be the best choice for our resolution analysis.

## 3.6    Resolution Treatment Method for 1D (Cylindrical Lens)

Assuming we have a cylindrical lens, we would like to know how we are able to apply the above resolution method for this type of lens. In previous section our focus was on microsphere lens which was an isotropic lens for 2D image along $x$ and $y$. In this section we take the asymmetric characteristics of cylindrical lens into account and modify our method for this lens.

Figure 3.2 (b) shows our general approach to study symmetric lens. The first step is drawing a simplified object with the perfect shape based on the results of SEM characterization. In our case, it represented by the circular dimers in Figure 3.2 (a). After that, we need to take into account that in contrast to spherical lens, the cylindrical lens has magnification across the fiber ($x$-axis) and it does not possess any magnification along the fiber ($y$-axis). Assuming the fiber is along the y axis and knowing that the image is magnified across the fiber, we adjusted our object for imaging by cylindrical lens by applying different magnifications along both $x$ and $y$ axis. Figure 3.5 (b) shows the modified object demonstrating the higher magnification across the fiber ($x$ axis).

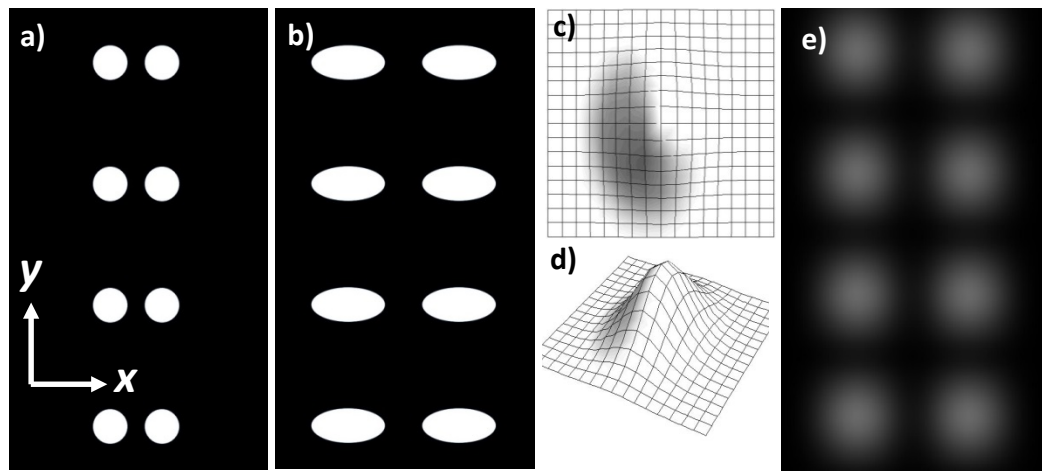

Figure 3.5: (a) Drawn trace of SEM image of the sample. (b) The modified trace by applying magnification across the direction of the fiber. (c), (d) The asymmetric Gaussian PSF from top view and 3D view respectively. e) The modified trace after being convolved with PSF.

The next step is to take into account that PSF can be also highly asymmetric in the case of cylindrical lens. In this case we assumed that the resolutions along the fiber and across the fiber are not the same which means FWHMs of the PSF along $x$ and $y$ is different and has to be considered separately. Figure 3.5 (c), (d) show a typical Gaussian PSF which has different FWHM along each axis. Such PSF can be used for convoluting with the object. By changing the PSF's width the image which resembles the object most closely can be identified. Figure 3.5 (e) shows a typical convolved image.

In order to fit the experimental profiles, after each convolution we have to compare the S/P ratio in both directions along the fiber and across the fiber. Figure 3.6 (a) is a calculated image of an array of dimers with 100 pixels diameters, 150 pixels separation along the $x$ direction and 350 pixels along the $y$ direction. The PSF has 125 pixels and 190 pixels width along the x and y directions respectively. However the separation along $x$ axis increases to 300 pixels as for the magnification.

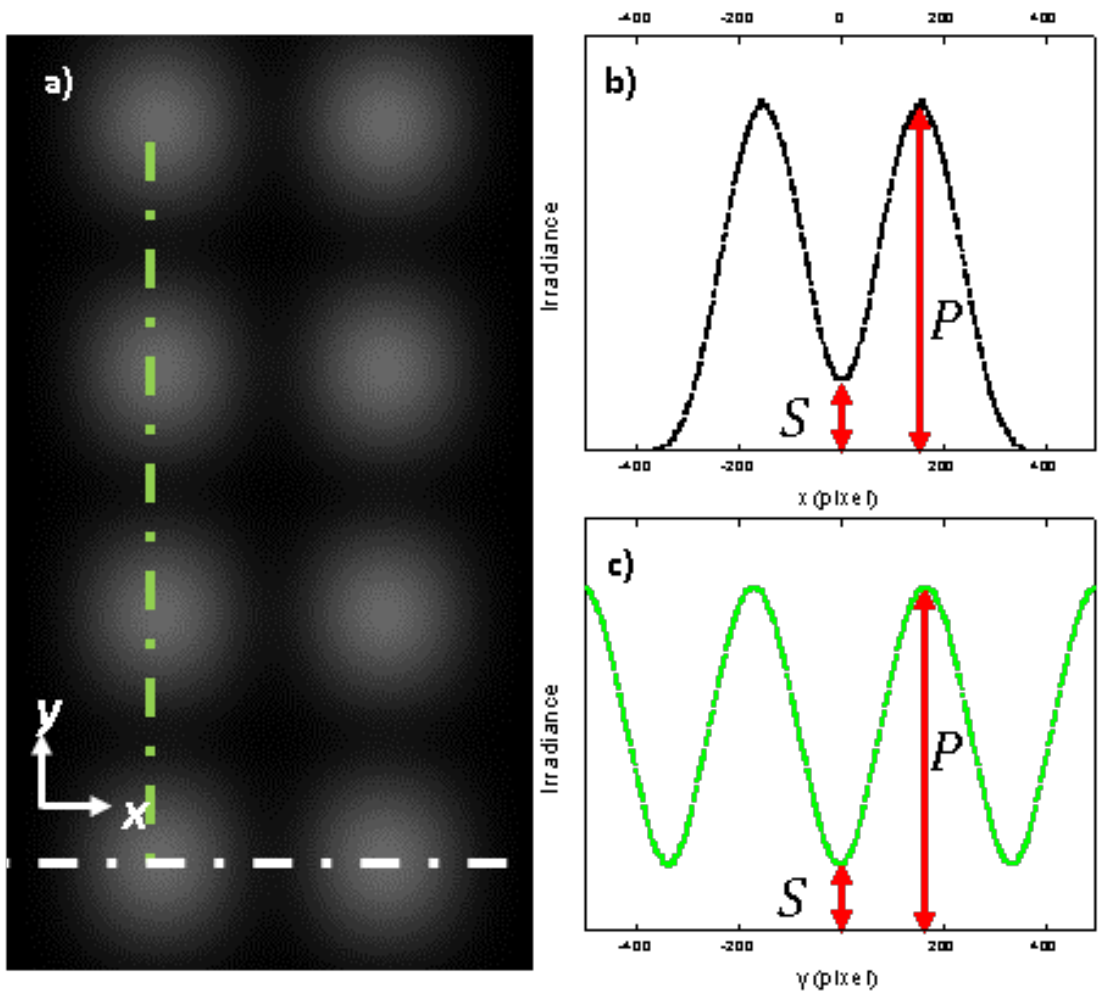

Figure 3.6: Analysis of the calculated intensity profile and demonstration of parameters saddle (*S*) and pick (*P*) (a) A typical calculated image of an array of dimers with 350 pixels separation along *y* axis and 150 pixels separation along *x* axis for a cylindrical lens (fiber) The separation across *x* direction increased to 300 pixels as for the magnification. (b) The calculated intensity profile along *x* direction (across the fiber) for the PSF with 125 pixels width. (c) The calculated intensity profile along y direction (along the fiber) for the PSF with 190 pixels width along the fiber.

Figure 3.6 (b) shows the typical calculated intensity profile across the fiber along *x* axis. We observed S/P ratio similar to what we previously presented in the case of imaging by microspheres, see Figure 3.3. It means we can use the PSF width and S/P ratio again as a useful parameter for comparison. Figure 3.6 (c) shows the typical intensity profile along the fiber. As it is noticeable, along the fiber we are able to see an array of dimers. There are many saddles and peaks which theoretically we can select for comparison with experiment. After all, the FWHM of the PSF which describes both S/P ratios along $x$ and $y$ determines the resolution in both directions.

3.7    Experimental Result and the Resolution by Using Our Model

In this section, we are going to apply all theoretical analysis we established above for a microsphere-assisted microscopy image. As Figure 3.7 (a) shows, the quality of this sample is low and the circular dimer during the fabrication process has become ellipse. But the advantage of this method is that regardless of the geometry of the structure this method can be employed to any well characterized sample. The image was taken through a BTG microsphere with the diameter of 8 µm. The dimer center-to-center separation is 180nm and the major and minor axes respectively are 140nm and 100nm.

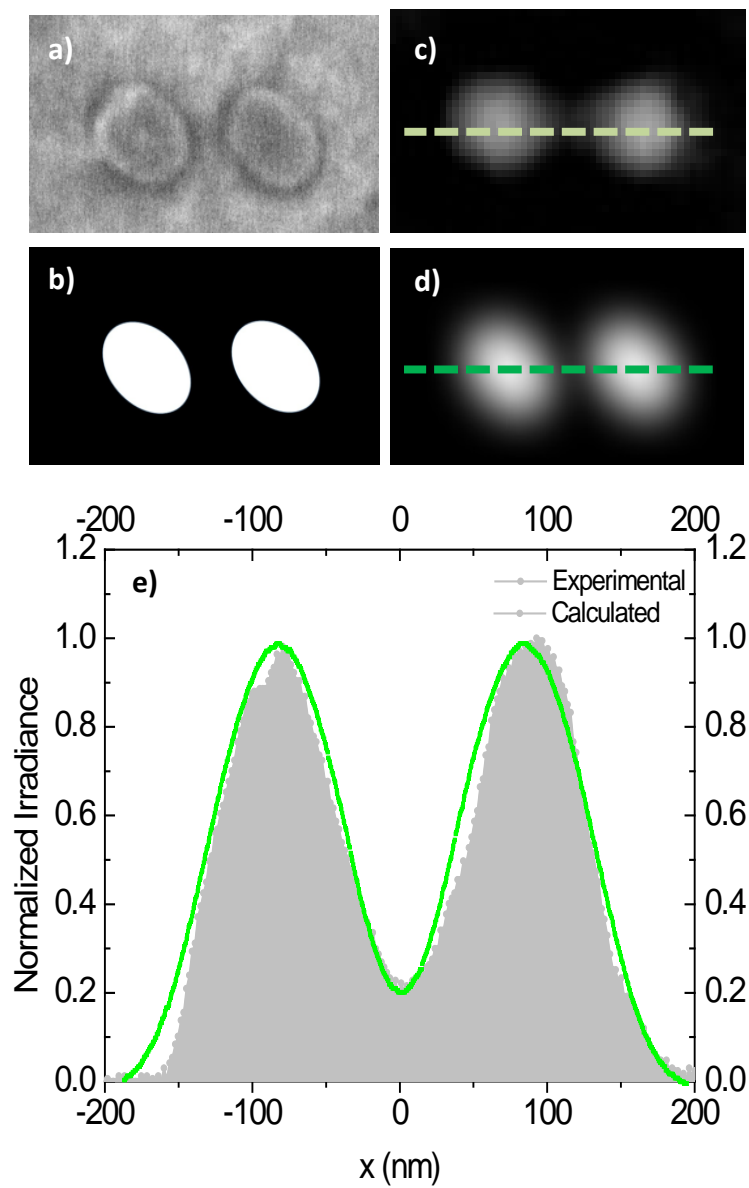

Figure 3.7: Experimental and modeling results. (a), (c) show the SEM and confocal microscope image of the aluminum sample. Dimer's major and minor diameters are 100nm and 140nm respectively and 180nm center-to-center separation. (b), (d) show the drawn (idealized) and calculated images. (e) Shows the experimental profile and the best fit.

The image was obtained using CLSM (LEXT OSL4000 with 405nm laser line). Figure 3.7 (b) shows the drawn idealized object based on the SEM characterization. (c and d) show respectively the confocal image of the dimer and the convolved image which is for the best fit by PSF ~70nm ($\lambda$/5.5).

The results presented in Figure. 3.7 show that generally our methodology of the resolution treatment works for images obtained using barium titanate spheres with diameter 8 µm. We believe that this method can be used with different spheres and different structures. Similar approach should be applicable to fitting images obtained by different super-resolution methods as well. The examples exclude near field scanning microscopy, far field super lenses, plasmonic nanogratings, two-photon absorption and other super-resolution techniques. The common approach to fitting images obtained by these techniques is based on using PSFs with the width which is not limited by the diffraction limit, as it is accepted in diffraction-limited optics.

Imaging by microspheres can be performed using spheres with different diameters. The results of these studies can be a subject of future work, however, we can hypothesize about the resolution expected for spheres with different diameters. In the limit of large, submillimeter-scale spheres we would expect the diffraction-limited resolution on the order of $\lambda/2n_o$. For spheres with $n$~2 this means that the resolution is limited at the level ~$\lambda$/4. For spheres with progressively smaller diameters the resolution can improve due to formation of photonic nanojets with very small sub-diffraction beam waists. We expect such behavior for spheres approximately within 3<D<10 µm range, however it requires more detailed experimental studies. For smaller spheres, the dielectric particles cannot operate as lenses, and they usually play the part of scattering particles. So, for such small spheres the imaging would not be possible.

## 3.8 Conclusion

The super-resolution imaging by films with embedded spheres is one of applications of so-called *microspherical photonics* actively developing in the Prof. Astratov's Lab. Microspherical photonics deals with the optical properties and applications of structures and devices formed by dielectric microspheres. The interest in this area is exponentially growing in recent years due to novel fundamental properties of individual and, especially, more complicated multisphere structures in some cases termed photonic clusters or molecules.

The terminology of "photonic atom" was first introduced by Arnold [50] in the context of understanding eigenmodes of spherical resonator based on an analogy quantum mechanical atoms. Mathematically, this analogy was pointed out by Nussenzveig [51] and Johnson [52]. Correspondingly, coupled photonic atoms can be termed "photonic molecules". In recent years this terminology has been used in a much broader context of cavities, not limited with microspheres. For example, photonic crystal cavities and semiconductor microcavities [53-56] can be also termed photonic atoms and corresponding coupled-cavity structures can be termed photonic molecules.

These include coupled whispering gallery modes [57-70], photonic nanojets and nanojet-induced modes [31, 71-82], resonant light pressure effects [8, 83-88], and optical super-resolution properties [32-34, 37, 89-92] Microspherical photonics can result in various applications such as sorting microspheres with almost identical resonant properties [8, 83-88], developing microsphere-chain waveguides [57-61, 31,82] and polarizers [79], surgical scalpels [72-78, 80, [93] and developing novel optical components for super-resolution imaging [35]. Three PhD dissertations were recently defended based on these studies in the Mesophotonics Lab at UNCC [4, 89, 94].

The imaging by microspheres emerged in 2011 [30]. However, the practical applications of this technology can be obtained by either immersing the high-index microspheres in a liquid, most importantly in water, or by embedding such spheres in elastomeric materials with ability to solidify. The water-immersion is critical for developing applications of this technology in life sciences where most of the cultures and samples are water-based solutions. On the other hand, immersion in elastomeric thin films, opens a way of developing surface scanning functionality of various solid-state and semiconductor samples and structures. Such solidified thin films containing embedded microspheres can be applied to investigated surfaces to achieve high-resolution microscopy. Both ways were invented in 2012 in the Mesoscale Photonics Lab by Prof. Vasily Astratov and his student, Arash Darafsheh [32].

In Chapter 2 of this MS thesis work, we developed a technology of embedding high-index BTG microspheres inside of transparent elastomeric PDMS slab. This PDMS slab can be considered as a novel optical component for super-resolution microscopy. Previously, embedding microspheres inside PDMS slabs have been used in different optical applications [?]. For super-resolution imaging applications, the important novel features were: i) using high-index barium titanate glass spheres and ii) positioning them in the optical near-field vicinity of the lower interface of the PDMS slab. The advantage of such thin films with embedded spheres is a possibility to have access to variety of microsphere sizes which provides different magnification and different resolution as well.

Another potential advantage is based on ability to shift the entire thin film with embedded spheres along the surface of investigated samples until some spheres would be aligned with the objects of studies. Such translation of thin films requires a liquid lubrication of the sample surface. Describing this functionality goes beyond the scope of this MS theses work, but generally it allows scanning large areas at the surface of investigated samples.

In Chapter 3 of this MS thesis work, we developed a rigorous method to determine the resolution of an arbitrary optical microscope, in our case microsphere-assisted microscope. Our approach is based on classical procedure of convolution with PSF well known for diffraction-limited systems. However, in our work we show that the same approach can be used for reconstructing the super-resolved images. The microscopic mechanism of super-resolution achieved in this method is debated in the literature. The possible mechanisms include so-called "photonic nanojet" properties [43, 44], excitation of resonant surface plasmons in metallic nanostructures, dielectric nanoantenna effect, or near-field excitation of whispering-gallery modes in dielectric microspheres. The study of specific mechanism goes beyond the scope of this work. We showed, however, that by using the PSF with subdiffraction-limited width, we can obtain images which are very similar to our experimental observations. We also performed the image analysis which shows the range of sizes of nanostructures which is most suitable for experimental studies of super-resolution. Using aluminum nanoplasmonic array we observed their virtual images through the microspheres and determined that the resolution casting on the order of $\lambda/6$ can be achieved in such cases.

We also developed our resolution analysis in the case of cylindrical lens formed by etched microfiber. We showed that it is possible to develop this method to calculate the super-resolution of a cylindrical lens by taking magnification across the fiber and the casymmetric PSF into account which both are imposed by the anisotropic structure of cylindrical lens.

For future work, having different nanoplasmonic samples, microsphere-embedded PDMS slab and this resolution treatment technique, we will be able to investigate the effect of Surface Plasmon Resonance (SPR) in super-resolution microsphere-assisted microscopy.

## APPENDIX: PEER-REVIEWED PUBLICATIONS